\documentclass[fleqn,usenatbib]{mnras}

\usepackage[T1]{fontenc}
\usepackage{ae,aecompl}

\usepackage[normalem]{ulem}
\usepackage{amsmath}
\usepackage{graphicx} 
\usepackage{lscape}
\usepackage{indentfirst}
\usepackage{enumitem}
\usepackage{xspace}

\usepackage{amssymb}
\usepackage{amsmath}
\usepackage{bm}
\usepackage[dvipsnames]{xcolor}
\usepackage{url}
\usepackage[flushleft]{threeparttable}

\newcommand {\ee}{\mathrm{e}}
\newcommand {\ii}{\mathrm{i}}
\newcommand {\beq}{\begin {equation}}
\newcommand {\eeq}{\end {equation}}

\newcommand{\dd}{\mathrm{d}}

\newcommand {\ergs}{\mbox{erg\ s$^{-1}$}}

\title[Magnetospheric flow ionization in XRPs]
{
Magnetospheric flows in X-ray pulsars II: \\
Heating, cooling and ionization degree at sub-critical accretion
}
\author[A. A.~Mushtukov et al.] 
{Alexander~A.~Mushtukov,$^{1,2}$\thanks{E-mail: alexander.mushtukov@physics.ox.ac.uk (AAM)}
Alexander~Y.~Potekhin,$^{3,4}$
Valery~F.~Suleimanov,$^5$
\newauthor
Andrew~Yu,$^{6}$
Sergey~S.~Tsygankov$^{7}$
\\ 
$^1$Mullard Space Science Laboratory, University College London, Holmbury St. Mary, Surrey RH5 6NT, UK\\
$^2$Astrophysics, Department of Physics, University of Oxford, Denys Wilkinson Building, Keble Road, Oxford OX1 3RH, UK \\ 
$^3$Ioffe Institute, Politekhnicheskaya 26, St Petersburg 194021, Russia \\
$^4$Space Research Institute of the Russian Academy of Sciences, Profsoyuznaya 84/32, Moscow  117997, Russia \\
$^5$Institut f\"{u}r Astronomie und Astrophysik, Kepler Center for Astro and Particle Physics, Universit\"{a}t T\"{u}bingen, \\ Sand 1, 72076 T\"{u}bingen, Germany\\
$^6$Department of Astronomy, Yale University, New Haven, CT 06520, USA  \\
$^7$Department of Physics and Astronomy,  FI-20014 University of Turku, Finland \\
}

\pubyear{2026}

\begin{document}
\label{firstpage}
\pagerange{\pageref{firstpage}--\pageref{lastpage}}
\maketitle

\begin{abstract} 
Magnetospheric accretion flows in X-ray pulsars shape their spectra, polarization, and variability. 
We model the thermal balance of the flow enveloping the neutron star magnetosphere in the sub-critical regime ($L \lesssim 10^{37}\,\mathrm{erg\,s^{-1}}$), where radiation forces do not control the dynamics and single Compton scatterings dominate. 
The energy budget includes Compton heating by surface X-rays, compressional (adiabatic) heating in the converging flow, and radiative cooling dominated by free-free emission and contributed also by cyclotron emission.
We show that the interplay of these processes leads to efficient cooling of the flow in the inner magnetosphere.
We compute the flow temperature profile as a function of luminosity and find that near the stellar surface the temperature can fall to a few tens of eV at $L < 10^{35}\,\mathrm{erg\,s^{-1}}$. 
Under such conditions, the accreting plasma, modelled here as pure hydrogen, is no longer fully ionized.
In the strong magnetic fields typical for X-ray pulsars, such temperatures permit partial recombination of electrons and protons into neutral hydrogen. 
As a result, a significant fraction of the flow becomes weakly ionized, while
external illumination ionizes this gas only partially within a geometrically thin layer immediately above the neutron star surface.
This implies that magnetospheric accretion at low luminosities proceeds through a partially ionized medium, in contrast to the commonly assumed fully ionized flow.
\end{abstract}


\section{Introduction}
\label{sec:Intro}

X-ray pulsars (XRPs) are accreting, strongly magnetized neutron stars (NSs) in close binary systems (see \citealt{2022arXiv220414185M} for a review). The magnetic field strength at the NS surface can reach up to $10^{13}\,\mathrm{G}$, as confirmed by the detection of cyclotron resonance scattering features in their spectra (see \citealt{2019A&A...622A..61S} for a review), and may be even stronger. Due to these extreme fields, XRPs serve as natural laboratories for studying fundamental physics under extreme conditions.

A magnetic field $B$ is considered strong when it exceeds the atomic unit of the magnetic field,  
\beq 
B_0 = \frac{m_\mathrm{e}^2e^3c}{\hbar^3} = 2.3505\times 10^9\,\mathrm{G},
\eeq 
where $m_\mathrm{e}$ is the electron mass, $e$ the elementary charge, $c$ the speed of light, and $\hbar$ the reduced Planck constant (using the Gaussian system of units). In such fields, the electron cyclotron energy,
\beq  
   E_\mathrm{cyc} = \frac{\hbar eB}{m_\mathrm{e} c}
   = 11.576\,B_{12}\text{~keV},
\eeq
where $B_{12} \equiv B/10^{12}$\,G,
exceeds 1 Hartree ($1\,\mathrm{Ha}=\alpha_\mathrm{f}^2 m_\mathrm{e}c^2 = 27.21$~eV, where $\alpha_\mathrm{f}$ is the fine structure constant), leading to a significant, non-perturbative modification of atomic structure.

For even stronger fields
\beq 
B_\mathrm{QED} = \frac{m_\mathrm{e}^2 c^2}{e\hbar} = 4.4140\times 10^{13}\,\mathrm{G},
\eeq 
the cyclotron energy surpasses the electron rest energy ($E_\mathrm{cyc} > m_\mathrm{e} c^2$), introducing significant quantum electrodynamics (QED) effects. The typical magnetic fields of XRPs lie in the regime where they are strong on the atomic scale but remain below the QED threshold ($B_0 \ll B \lesssim B_\mathrm{QED}$).  

The accretion luminosity at the surface of an XRP is often expressed in terms of the mass accretion rate $\dot{M}$, neglecting General Relativity (GR) effects, as  
\beq 
L_\mathrm{N}= \frac{G\dot{M}M}{R} = 1.858\times 10^{37}\,\dot{M}_{17}\frac{M_{1.4}}{R_6}\,\,\ergs,
\label{L_Newton}
\eeq 
where $\dot{M}_{17} \equiv \dot{M}/10^{17}$  g\,s$^{-1}$, $M_{1.4} = M/1.4\,M_\odot$, $R_6 = R/10^6\,\mathrm{cm}$, $M$ being the NS mass, $M_\odot$ the solar mass, $R$ the NS radius and $G$ is the Newtonian gravitational constant.

When GR effects are considered, the bolometric accretion luminosity $L$ measured at infinity is related to the local accretion rate at the NS surface by \citep[e.g.,][]{Mitra98,Meisel_18}  
\beq
   L = \frac{z}{(1+z)^2}\,\dot{M}c^2 = \frac{L_\mathrm{N}}{1+z/2},
\label{LXvsMdot}
\eeq
where the gravitational redshift at the stellar surface is  
\beq
   z = \frac{1}{\sqrt{1 - r_\mathrm{S}/R}} - 1,  
\eeq
with $r_\mathrm{S} = 2GM/c^2 = 4.134\,M_{1.4}$ km being the Schwarzschild radius. Here, $L_\mathrm{N}$ is the Newtonian accretion luminosity given by equation~(\ref{L_Newton})
and $\dot{M}$ is measured in the local reference frame at the NS surface. 

Most of the energy is typically released in X-ray energy band. 
Apparent luminosity of XRPs covers a wide range from $\sim 10^{32}$ \ergs\ up to $~10^{41}$ \ergs, where the brightest objects belong to the class of pulsating ultraluminous X-ray sources (ULXs, see \citealt{2021AstBu..76....6F,2023NewAR..9601672K} for review). 
The lower luminosity limit is related to (i) the sensitivity of X-ray missions and (ii) appearance of specific states of accretion: the propeller effect, when rotating magnetic field of a NS sets up a centrifugal barrier and prevents stable accretion onto NS surface with observed emission arising from the cooling of the NS \citep{1975A&A....39..185I,2006ApJ...646..304U,2016A&A...593A..16T,2025arXiv250909860X}, or stable accretion from cold disc of low viscosity \citep{2017A&A...608A..17T,2019A&A...621A.134T}.
Appearance of the propeller state results in a sharp drop of pulsar X-ray luminosity to the level of thermal
flux
from the polar caps as soon as accretion luminosity is below the 
propeller transition luminosity
\citep{2002ApJ...580..389C}
\beq 
L_\mathrm{prop}\simeq 4 \times 10^{37} \Lambda^{7/2} B_\mathrm{p12}^2
P_\mathrm{s}^{-7/3} M_{1.4}^{-2/3} R_6^5\,\,\ergs,
\eeq 
where $B_\mathrm{p12}=B_\mathrm{p}/10^{12}$~G, $B_\mathrm{p}$ is the magnetic field strength at the pole, $P_\mathrm{s}$ is the spin period in seconds and $\Lambda$ is a coefficient related to the geometry of accretion flow, that is taken to be $\sim 0.5$ for accretion from a geometrically thin disc.
Appearance of a cold disc on the contrary prevents
a decrease
of X-ray luminosity below \citep{2017A&A...608A..17T}
\beq
L_\mathrm{cold} \simeq 9\times 10^{33}\,\Lambda^{1.5}B_\mathrm{p12}^{0.86}M_{1.4}^{0.28}R_6^{1.57}\,\,\ergs, 
\eeq 
and an XRP can stay in this state for a long period of time while
its accretion disc slowly loses material due to low level accretion. 

In addition to these scenarios, an alternative regime of accretion at low luminosities has been proposed in the form of quasi-spherical settling
\citep{2012MNRAS.420..216S,2015ARep...59..645S}, where a hot, quasi-static envelope is formed around the NS magnetosphere. 
In this case, the physical conditions of the plasma interacting with the magnetosphere may differ from those in disc-fed accretion.
A detailed treatment of this regime is beyond the scope of this paper.

An accretion flow from a companion star in XRPs is interrupted by a strong magnetic field at a certain distance from a NS called magnetospheric radius
(about one half of the Alfv\'en radius; cf.{} equation (6.20) in \citealt*{2002apa..book.....F}):
\beq\label{eq:Rm}
R_\mathrm{m} \simeq 1.7\times 10^8\,\Lambda B_\mathrm{p12}^{4/7}\dot{M}_{17}^{-2/7}M_{1.4}^{-1/7}R_6^{12/7}\,\,\mathrm{cm}.
\eeq
Within the magnetospheric radius, a strong magnetic field of a NS shapes the geometry of an accretion flow: it is expected that hot ionized plasma has to follow magnetic field lines towards the surface of a NS and accrete onto the stellar surface within small regions of area $\sim 10^{9}\,\mathrm{cm^2}$.

At higher accretion luminosities ($>10^{38}$ \ergs), the flow within the magnetosphere of a NS becomes optically thick due to Compton scattering 
\citep[e.g.,][]{2017MNRAS.467.1202M}. 
Under these conditions, the magnetospheric accretion flow significantly influences X-ray spectra, pulse profiles, and the timing features of X-ray aperiodic variability \citep{2019MNRAS.484..687M,2023MNRAS.525.4176B}. As the flow progresses toward the NS surface, it can undergo cooling and heating through free-free absorption/emission, Compton scattering, and compression \citep{1976SvAL....2..111S}.
The temperature and ionization state of the accretion flow within the magnetosphere and at the NS surface can directly impact plasma braking processes in close proximity to the stellar surface and affect polarization formation, even at low mass accretion rates \citep{2022ApJ...941L..14T}. Despite these implications, the details of the plasma heating and cooling processes in the magnetospheres of XRPs have not yet been quantitatively analysed.

In this paper, we investigate the dynamics of the accretion flow within the magnetosphere of a NS dominated by a dipole magnetic field. We focus on variations in the temperature of the accretion flow, taking into account free-free emission and Compton scattering of X-ray photons. Our analysis is restricted to low mass accretion rates, corresponding to luminosities $\lesssim 10^{37}$ \ergs, where the accretion flow remains optically thin.

\begin{figure}
\centering 
\includegraphics[width=\columnwidth]{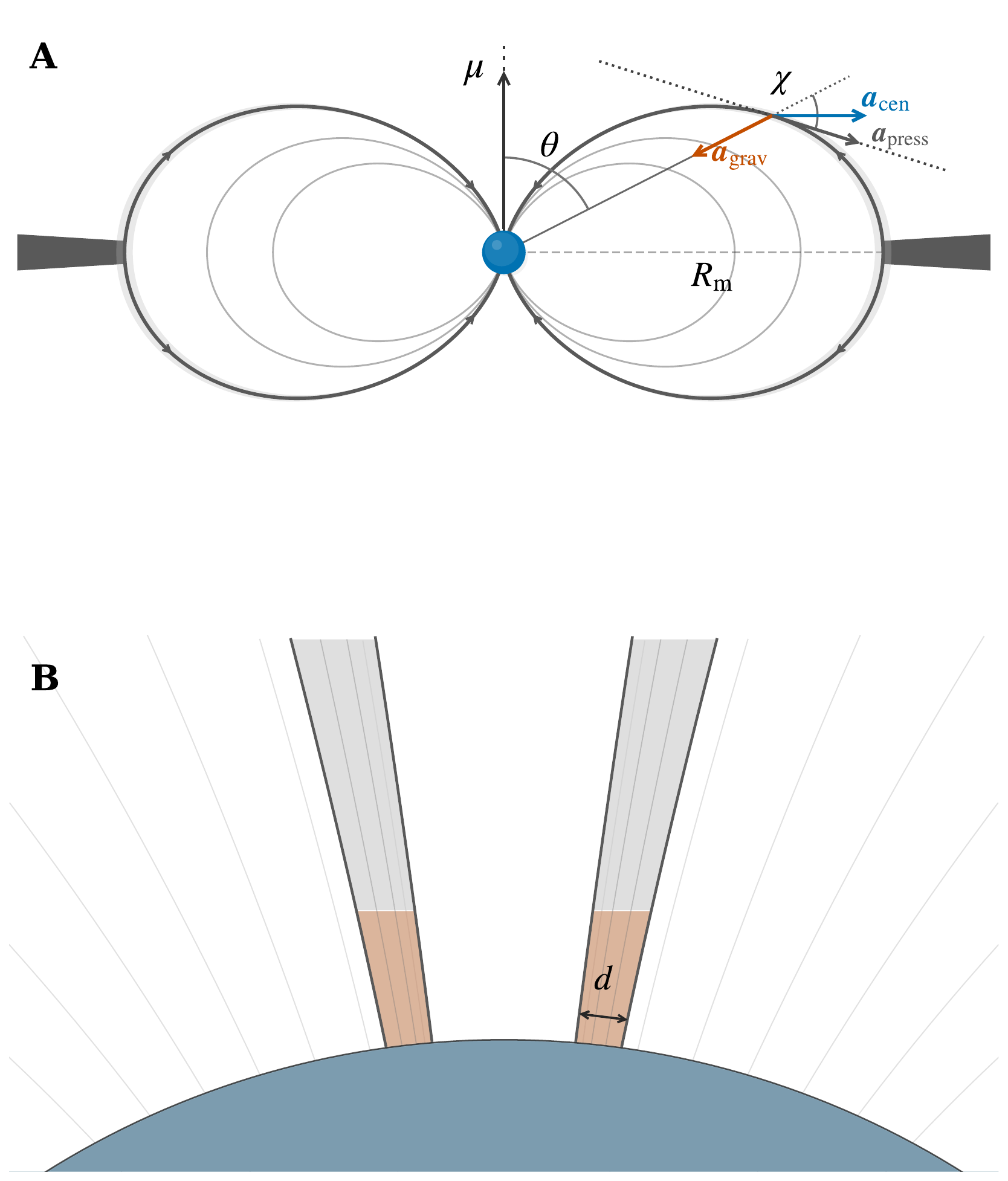}
\caption{
The schematic representation of the dipole magnetic field lines, where $R_\mathrm{m}$ denotes the inner disc radius, $\theta$ is the magnetic colatitude of a given point and $\chi$ is the angle between the position vector of this point and the tangent to the field line (\ref{eq:chi}). \\
Panel A illustrates the global geometry of the magnetosphere and the main forces acting on the accretion flow, including gravitational, centrifugal and pressure-gradient accelerations.\\
Panel B shows a zoomed-in view of the accretion channel near the neutron star surface, which defines the geometry adopted in our calculations. In this region, the flow may become partially recombined at low luminosities, while external X-ray illumination produces a geometrically thin photoionized layer located close to the NS surface.
The shaded grey region in panel B schematically marks the partially recombined part of the accretion flow, while the layer adjacent to the NS surface (colored in brown) shows the photoionized layer produced by X-ray illumination.
}
\label{pic:dipole_scheme}
\end{figure}

\section{Model}
\label{sec:Model}

We consider a NS  accreting from a geometrically thin disc. 
The NS magnetic field is assumed to be dominated by its dipole component. 
The rotational axis is taken to be perpendicular to the plane of the accretion disc, while the magnetic dipole moment $\bm{\mu}$ may be inclined relative to the rotational axis. 

For simplicity, we adopt a non-relativistic dipole model, neglecting distortions of the magnetic field due to GR effects \citep{GinzburgOzernoi}. 
The accretion flow originates from the disc plane at the magnetospheric radius, $R_\mathrm{m}$ (\ref{eq:Rm}), with an initial velocity comparable to the local thermal velocity of protons. 
Within the magnetosphere, the flow is assumed to be strictly channelled along the magnetic field lines,
being primarily influenced by gravitational, centrifugal, Coriolis forces, and gas pressure-gradient acceleration, while the radiative force is assumed to be negligible.

The thermal evolution of the accretion flow is governed by three key processes: free-free emission, cyclotron emission, Compton scattering of photons from the NS surface, and compressional heating as the flow approaches the star.  

The dynamical structure of the magnetospheric flow adopted in this work follows our previous studies \citep{2024MNRAS.530..730M}. 
In this framework, the accretion flow is assumed to be confined to dipole magnetic field lines and its motion is governed by gravitational, centrifugal, Coriolis and pressure-gradient forces, while radiative forces are neglected in the sub-critical regime considered here. 
The resulting velocity and density profiles along the field lines are then used as input for the thermal and ionization calculations presented below.

\subsection{Geometry {and dynamics} of accretion flow}

\subsubsection{Dipole magnetic field geometry}
\label{sect:geom}

The inclination of the magnetic field axis relative to the rotation axis is defined by the angle \(\alpha \in [0, \pi/2]\). 
Using spherical coordinates \((r, \theta, \varphi)\), where \(\theta = 0\) corresponds to the direction of the magnetic field axis,
a dipole magnetic field line is given by
\begin{equation}
\label{eq:dip_fl}
r(\theta, \varphi) = R_\mathrm{max}\sin^2 \theta,
\end{equation}  
where \(R_\mathrm{max}\) represents a linear scale specific to a particular magnetic field line. It is expressed as
\begin{equation}
R_\mathrm{max} = R_\mathrm{m}/\sin^2\theta_\mathrm{d},
\end{equation}  
with the inner disc radius
$R_\mathrm{m}$ given by equation~(\ref{eq:Rm}) 
and 
\begin{equation}
\theta_\mathrm{d} = \arctan([\sin\varphi \, \tan\alpha]^{-1}),
\end{equation}  
describing the angular displacement of the field line.
Local strength of magnetic field is given by
\beq \label{eq:B}
B \simeq \frac{B_\mathrm{p}}{2}\left(\frac{R}{r}\right)^3\,\sqrt{1+3\cos^2\theta}.
\eeq
In this paper we restrict ourselves by the particular case where $\alpha=0$, which is depicted in Fig.~\ref{pic:dipole_scheme}A. In this case, $\theta_\mathrm{d}=\pi/2$ and $R_\mathrm{max}=R_\mathrm{m}$.

The linear distance $s$ along a magnetic field line relates to the angular coordinate \(\theta\) through
\begin{equation}
\label{eq:dx}
\dd s = R_\mathrm{max}\sin\theta \sqrt{1 + 3\cos^2\theta} \, \dd\theta.
\end{equation}  
The corresponding elementary area at the magnetospheric surface is 
\begin{equation}
\dd S = R_\mathrm{max}^2 \sin^4\theta \sqrt{1 + 3\cos^2\theta} \, \dd\theta \, \dd\varphi.
\end{equation} 
The cross-sectional area of an individual magnetic flux tube will be denoted by \(A(s)\), where \(s\) is the distance measured along the magnetic field line. 
Magnetic flux conservation implies
\begin{equation}
    A(s) B(s) = {\rm const},
\end{equation}
so that \(A\propto B^{-1}\) along the flow.

\subsubsection{Weak and strong field regimes}
\label{sec:weak_strong}

As mentioned in the Introduction, the strong magnetic field $B\gg B_0$ significantly modifies quantum-mechanical atomic properties. In addition, if the typical kinetic energies of the plasma particles are not large compared with the electron cyclotron energy,
then the Landau quantization of the electron motion across the field lines is no longer smeared away by thermal averaging and should be taken into account.
In a non-degenerate plasma, the latter condition means $T \lesssim E_\mathrm{cyc}$,
where $T$ is the temperature in energy units.
Under these conditions, the cross sections of interactions of the electrons with other plasma particles (protons and atoms) and with photons (in particular, Compton scattering cross sections) strongly differ from the non-magnetic ones (see, e.g., \citealt{HardingLai06}, and references therein).

In the dipole geometry of Section~\ref{sect:geom}, the condition $B > B_0$ is fulfilled at the distances $r < R_*$ from the NS centre and the condition $T < E_\mathrm{cyc}$ is fulfilled at $r < R'_*$, where 
$R_* \approx 7.5 R B_\mathrm{p12}^{1/3}$, $R'_* \approx 5 R (B_\mathrm{p12}/T_{100})^{1/3}$ and $T_{100} \equiv T / 100$~eV.
These estimates follow directly from the dipole scaling, equation~(\ref{eq:B}), by setting \(B=B_0\) and \(E_{\rm cyc}=T\), respectively, and keeping only order-of-magnitude numerical factors.
For the values of $B_\mathrm{p}$ and $T$ encountered below, both $R_*$ and $R_*'$ are of the order of $10R$.
Therefore, for order-of-magnitude estimates, one may use the non-magnetic formulae for plasma thermodynamics and cross sections of processes involving free or bound electrons in the outer part of the magnetosphere, where $r\gtrsim 10R$, and the formulae valid in strong (and strongly quantizing) magnetic field in the inner part, at $r\lesssim10R$. Hereinafter we will briefly call these alternative cases as the weak and strong field regimes, respectively.

One can treat electrons as confined to the ground Landau level and protons as classical particles, if $T$ is small compared with $E_\mathrm{cyc}$ but larger than the proton cyclotron energy $E_\mathrm{cyc,p} = E_\mathrm{cyc}m_\mathrm{e}/m_\mathrm{p}$,
that is
\beq
0.063 \ll \frac{T_{100}}{B_{12}} \ll 116.
\eeq
We shall use such simplified treatment for making rough estimates in the strong-field regime, although the left inequality can be violated at the highest $B$ and lowest $T$ considered below. However, in numerical calculations of the ionization degree  (Section~\ref{sec:Ionization_degree}) we take an accurate account of the Landau quantization for protons.


\subsubsection{Gravitational, centrifugal and Coriolis accelerations}

In the regime of relatively low mass accretion rates and luminosities in XRPs in the reference frame co-rotating with an NS, the plasma motion along magnetic field lines is governed by gravitational, centrifugal and Coriolis forces.
Neglecting the GR factor $(1-r_\mathrm{S}/r)^{-1/2} \simeq (1-0.004M_{1.4}/r_8)^{-1/2}$,
the gravitational acceleration along the magnetic field lines is
\begin{equation}\label{eq:a_grav}
\left|a_{\mathrm{grav},||}\right| = \cos\chi\,\frac{GM}{r^2} = 1.858 \times 10^{10}\, M_{1.4} r_8^{-2} \cos\chi \mbox{ cm\,s$^{-2}$},
\end{equation}  
where \(\chi\) is the angle between the position vector of a point on the NS magnetosphere and the tangent to the field line (see Fig.\,\ref{pic:dipole_scheme}A), given by
\begin{equation}\label{eq:chi}
\chi = \arctan(0.5 \, \mathrm{tan}\theta).
\end{equation}  

The centrifugal acceleration along the field lines is
\begin{equation}\label{eq:a_cent_gen}
a_{\mathrm{cen},||} = -\bm{n}_B \cdot [\bm{\omega} \times (\bm{\omega} \times \bm{r})],
\end{equation}  
where \(\bm{n}_B\) is the unit vector tangent to the magnetic field line, \(\bm{\omega}\) is the angular velocity vector, and \(\bm{r}\) is the position vector.

The plasma moving along magnetic field lines in the rotating frame is subject to the Coriolis acceleration, which has a component along the field line given by
\begin{equation}\label{eq:a_cor}
a_{\mathrm{cor},||} = -2\,\bm{n}_B \cdot (\bm{\omega} \times \bm{v}),
\end{equation}
where \(\bm{v}\) is the plasma velocity in the co-rotating frame. 
The Coriolis term becomes significant when the velocity along the field lines is large.
In terms of characteristic values,
\begin{equation}
{|a_{\mathrm{cor},||}| \simeq 1.3 \times 10^{9}\, P_\mathrm{s}^{-1}\, v_9 \cos\zeta \;\; \mathrm{cm\,s^{-2}},}
\end{equation}
where
\(v_9 = v/(10^9\,\mathrm{cm~s^{-1}})\)
and \(\zeta\) is the angle between the field-line tangent and the Coriolis acceleration vector.

\subsubsection{Gas pressure gradient acceleration}

In addition to gravitational, centrifugal, and Coriolis forces, the plasma experiences acceleration due to the gradient of gas pressure. 
The corresponding acceleration along magnetic field lines can be written as
\begin{equation}
a_{\mathrm{press},||} = - \frac{1}{\rho} \frac{\dd P}{\dd s},
\end{equation}
where $s$ is the coordinate along the magnetic field line, $P$ is the gas pressure, 
 $\rho$ is the local mass density.
This term describes acceleration (or deceleration) of the plasma caused by
variations in temperature and density along the field lines.

In the following we will consider the partially ionized hydrogen plasma with ionization degree $f_+ = 1 - f_\mathrm{H} = n_\mathrm{p}/n_\mathrm{p,tot}$, where $f_\mathrm{H} = n_\mathrm{H}/n_\mathrm{p,tot}$ is the neutral fraction, $n_\mathrm{p}$ is the number density of free protons, equal to the number density of free electrons $n_\mathrm{e}$ due to the charge neutrality condition, $n_\mathrm{H}$ is the number density of H atoms, $n_\mathrm{p,tot} \approx \rho/m_\mathrm{p}$ is the total number density of protons (free and bound)
 and $m_\mathrm{p}$ is the proton mass.
Neglecting the molecules and using the ideal-gas equation of state for simplicity, we have $P=(n_\mathrm{p}+n_\mathrm{e}+n_\mathrm{H})\,T = n_\mathrm{p,tot}(1+f_+)\,T$. 

Assuming that the mass flux along the field line is conserved,
\(\dot M = \rho v A = {\rm const}\), and using magnetic-flux conservation,
\(A\propto B^{-1}\), the derivative of the density logarithm over the path length can be expressed as
\begin{equation}
\frac{1}{\rho}\frac{d\rho}{\dd s} = - \frac{1}{v}\frac{\dd v}{\dd s} + \frac{1}{B}\frac{\dd B}{\dd s}.
\end{equation}

Therefore, the pressure-gradient acceleration takes the form
\begin{equation}
\label{eq:a_press}
a_{\mathrm{press},||} = - \frac{T}{m_\mathrm{p}} 
\left(
  \frac{1}{T}\frac{\dd T}{\dd s}
  - \frac{1}{v}\frac{\dd v}{\dd s}
  + \frac{1}{B}\frac{\dd B}{\dd s}
  + \frac{1}{1+f_+}\,\frac{\dd f_+}{\dd s}
\right).
\end{equation}
This expression allows one to evaluate the local contribution of the gas-pressure gradient once the temperature, velocity and ionization degree profiles $T(s)$, $v(s)$ and $f_+(s)$ are known (see Section~\ref{sec:NumMod}).

For an order-of-magnitude estimate, taking a characteristic temperature $T = 0.1$~keV and a pressure scale length $l = 10^{8}\text{ cm}$, one finds
\begin{equation}
|a_{\mathrm{press},||}| 
\sim \frac{T}{m_\mathrm{p} l} 
\simeq 8\times 10^5\,T_{100}\,l_{8}^{-1}\mathrm{~cm~s^{-2}},
\end{equation}
where $T_{100} = T/100$~eV and $l_{8} = l/(10^{8}\,\mathrm{cm})$. 
This acceleration is typically one to two orders of magnitude smaller than the 
gravitational acceleration, but it can become comparable to or exceed the Coriolis term in the outer parts of the magnetosphere, especially when steep temperature gradients arise due to Compton heating or radiative cooling.

\subsection{Cooling and heating of accretion flow}

Cooling and heating of magnetospheric accretion flow is determined by free-free emission of hot plasma and energy exchange due to Compton scattering of X-ray photons originated from the surface of a NS. 
The temperature profile is obtained by solving the energy equation along the flow.
We describe the flow locally by its geometrical thickness \(H\) across the magnetic field lines and by the surface density
\begin{equation}
    \Sigma = \rho H ,
\end{equation}
where \(\rho\) is the mass density. In our formulation, the energy equation can be written as
\beq
\label{eq:T_evolution}
\frac{\dd T}{\dd t}
=
\frac{2 m_{\mathrm{p}}}{3}
\left(
\frac{Q^{+}-Q^{-}}{\Sigma}
\right),
\eeq
where $\dd/\dd t \equiv v\,\dd/\dd s$ is the derivative along the flow, $\Sigma$ is the local surface density
of mass, 
and $Q^{+}$ and $Q^{-}$ are the total heating and cooling rates per unit surface area. 
In particular, we include the contributions $Q^{+} = Q^{+}_{\rm sc} + Q^{+}_{\rm comp}$ and $Q^{-} = Q^{-}_{\rm ff} + Q^{-}_{\rm cyc}$,
where \(Q^+_{\rm sc}\) is Compton heating and \(Q^+_{\rm comp}\) is compressional heating.

We approximate the initial angular distribution of X-ray luminosity by 
\beq \label{eq:beam}
\frac{\dd L_\mathrm{i}}{\dd\Omega}\propto 
|\cos^\xi \theta_{B}|,
\eeq 
where $\theta_{B}\in[0,\pi]$ is the polar angle in the reference frame related to the magnetic dipole and $\xi>0$ is a parameter.
To describe spectral distribution of X-ray photons emitted at magnetic poles of a star, we apply
either the black-body spectrum of a given temperature $T_\mathrm{sp}$ or a
power law spectrum with an exponential cut off.
Rough estimations performed for the case of accretion from the disc (see equation 17 in \citealt{2015MNRAS.447.1847M}) show that the effective temperature of hot spots at the NS surface is
\beq\label{eq:T_eff_sp}
T_\mathrm{sp}\approx 7\,\Lambda^{7/32} B_\mathrm{p12}^{1/8}L_{37}^{3/20} 
M_{1.4}^{13/80} R_6^{-19/40}\,\,\mathrm{keV},
\eeq
where $L_{37} = L/10^{37}$ erg s$^{-1}$.

\subsubsection{Free-free and cyclotron emission}
\label{sec:free-free_emission}

The energy losses of hot ionized non-magnetized plasma due to free-free emission can be roughly estimated as 
\beq\label{eq:Q_min_1}
Q^{-}_\mathrm{ff}\simeq 
1.5\times 10^{14}\,T_{100}^{1/2}
\,n_\mathrm{e,19} n_\mathrm{i,19} Z^2\overline{g}_B~\mbox{ erg cm$^{-3}$\,s$^{-1}$},
\eeq 
where $n_\mathrm{e,19} \equiv n_\mathrm{e}/(10^{19}$ cm$^{-3}$) and  $n_\mathrm{i,19} \equiv n_\mathrm{i}/(10^{19}$ cm$^{-3}$),
$n_\mathrm{e}$ and $n_\mathrm{i}$ being the number densities of free electrons and ions respectively, $Z$ is the charge number of ions, and $1.1\lesssim\overline{g}_B\lesssim 1.5$ is the frequency and velocity averaged Gaunt factor (see Chapter 5 in \citealt{1986rpa..book.....R};
for numerical estimates
we choose $\overline{g}_B = 1.2$).  
Note that when considering plasma with solar chemical composition or close to it, the cooling in spectral lines can be significant, (see, e.g., \citealt{2009A&A...508..751S}). Therefore, equation  (\ref{eq:Q_min_1}) can be considered as a rough estimate of the cooling function from below.
Since
it is non-magnetic, it is applicable only
in the weak-field domain.

For the hydrogen plasma, we replace notation $n_\mathrm{i}$ by $n_\mathrm{p}$ and set $Z=1$. Then $n_\mathrm{e}=n_\mathrm{p}=n_\mathrm{p,tot}f_+$,
$n_\mathrm{e,19} \approx 0.6\, \rho/(10^{-5}$ g cm$^{-3}$),
and the energy losses due to free-free emission from the unit area of an optically thin magnetospheric accretion flow can be estimated as
\beq\label{eq:Q_min_H}
Q^{-}_\mathrm{ff}H\simeq 1.76\times 
10^{23}T_{100}
\Sigma_1^2 f_+^2 \overline{g}_B H_1^{-1} \,\,\,\mbox{ erg~cm$^{-2}$\,s$^{-1}$},
\eeq 
where \(\Sigma_1 = \Sigma/(1\,{\rm g\,cm^{-2}})\)
and
\(H_1 \equiv H/(1\,{\rm cm})\).

The energy losses due to the cyclotron emission for the case of optically thin plasma and $T\gg E_\mathrm{cyc}$ can be estimated as
\begin{align}&
Q^{-}_\mathrm{cyc} = \sigma_\mathrm{T}c n_\mathrm{e}\frac{B^2}{2\pi}\frac{T_\mathrm{e}}{m_\mathrm{e}c^2}
\nonumber\\&\qquad
\simeq 6.2\times 
10^{12}\,B_6^2\, T_{100}\, 
n_\mathrm{e,19}\,\,\,\,\mathrm{erg\,cm^{-3}\,s^{-1}},
\end{align}
where $\sigma_\mathrm{T}$ is the Thomson cross section,
$B_6\equiv B/10^6$~G (see, e.g., Chapter~5 in \citealt*{1995pprc.book.....D}).
However, the magnetospheric accretion flow can be optically thick at the fundamental electron cyclotron resonance and its harmonics \citep*{1980Ap&SS..73...33P}.
The corresponding energy loss from unit area is
\beq
Q^{-}_\mathrm{cyc}H \simeq 3.7\times 
10^{17}\, T_{100}
\,\Sigma f_+ B_6^2 \,\,\,\mbox{ erg\,cm$^{-2}$\,s$^{-1}$}.
\eeq
Assuming that the penetration depth of accretion disc into the magnetosphere is close to its geometrical thickness $H_\mathrm{d}$ at the inner radius, one can estimate the geometrical thickness of the flow as
\beq 
H \approx H_\mathrm{d}\sin^2\theta \sin\chi.
\eeq 
In the case of relatively low mass accretion rates, the disc is interrupted in the so-called $C$-zone,
where the pressure is dominated by gas pressure and the opacity is shaped by free-free absorption \citep*{2007ARep...51..549S}.
Then the geometrical thickness at the magnetospheric radius can be estimated as
\begin{align}&
H_\mathrm{d} \approx 1.2\times 10^{6}\,\alpha_\mathrm{d}^{-0.1}L_{37}^{3/20}M_{1.4}^{-21/40}R_6^{3/20}R_\mathrm{m,8}^{9/8}\,\,\,\mathrm{cm} 
\\&\qquad
\simeq
2.2\times 10^6\,\Lambda^{9/8} B_\mathrm{p12}^{9/14}\dot{M}_{17}^{-6/35}\alpha_\mathrm{d}^{-0.1}M_{1.4}^{-15/28}R_6^{27/14}\,\,\mathrm{cm},
\nonumber
\end{align}
{where $R_\mathrm{m,8} = R_\mathrm{m}/10^8$\,cm}
and $\alpha_\mathrm{d}<1$ is a dimensionless viscosity parameter of accretion disc \citep{1973A&A....24..337S}.


The optical thickness of the accretion channel at the fundamental electron cyclotron resonance and its harmonics is evaluated using the expressions for the cyclotron opacity given in Appendix~\ref{app:CyclotronOpacity}. 
In the original estimate, when the flow was optically thick at a given harmonic, the corresponding radiative loss was limited by the blackbody photon flux. 
However, in the strongly quantizing regime this prescription should be regarded only as an upper-limit estimate to the true cyclotron photon production rate. 
The reason is that resonant cyclotron interactions are dominated by scattering, while true photon production is controlled by magnetic free-free transitions with resonant
structure, as discussed in Appendix~\ref{sec:Magnetic_free-free_cooling_new}.

To assess the impact of this uncertainty on the thermal structure, we performed an additional calculation in which the cyclotron cooling term was artificially removed, \(Q^{-}_{\rm cyc}=0\). 
The results are presented in Appendix~\ref{sec:Magnetic_free-free_cooling_new}.
The temperature in the strong-field zone near the surface, where the recombination is important,
is practically unchanged. 
Therefore the main results of this work do not rely on the approximate LTE treatment of cyclotron cooling.

\subsubsection{Compton scattering}
\label{sec:Compton}

Compton scattering of X-ray photons produced at the NS surface by the electrons of an accretion flow can heat up or cool down the accretion flow depending on the flow temperature, velocity, energy and initial direction of X-ray photons. 
Calculation of heating/cooling due to the Compton scattering, we assume that electrons are moving strictly along magnetic field lines, and their distribution over the momentum along magnetic field lines $p_z$ is determined by a local velocity
$v$ and local temperature $T$ of the flow according to
the one-dimensional Maxwell--J\"uttner distribution \citep*[e.g.,][]{2010PhRvE..81b1126C}, which can be written in the NS rest frame as
\beq
\label{eq:Maxwell_moving_electrons}
f_\mathrm{e,M}(p_z,T,v)
= \frac{\ee^{-y[\gamma_0(\gamma-\beta_0 \hat{p}_z)]}}{2\gamma_0 K_1(y)},
\eeq
where $\beta_0\equiv v/c$, $\hat{p}_z \equiv p_z/m_\mathrm{e} c$, $\gamma\equiv\sqrt{1+\hat{p}_z^2}$ and $\gamma_0=(1-\beta_0^2)^{-1/2}$ are the dimensionless velocity of the flow, dimensionless momentum of an electron and the Lorentz factors of the electron and the accretion flow, respectively, $y\equiv m_\mathrm{e}c^2/T$ and $K_{1}(y)$ is the modified Bessel function of the second kind. 
The electron distribution function (\ref{eq:Maxwell_moving_electrons}) is normalized as 
\beq
\int_{-\infty}^{\infty}\dd \hat{p}_z\,f_\mathrm{e,M}(p_z,T,\beta_0)=1.
\eeq
In the case of non-relativistic temperatures of the flow ($y\gg 1$), equation~(\ref{eq:Maxwell_moving_electrons}) can be simplified to 
\beq\label{eq:Maxwell_moving_electrons_app}
f_\mathrm{e,M}(p_z,T,\beta_0)
= \gamma_0^{-1}\sqrt{\frac{y}{2\pi}}\ee^{y[1-\gamma_0(\gamma-\beta_0 \hat{p}_z)]}.
\eeq 

In the weak-field regime, we describe Compton scattering in non-magnetic approximation.
Then the differential cross section in the electron's rest frame is given by Klein-Nishina formula
\beq\label{eq:cross_section_diff}
\frac{\dd\sigma}{\dd\Omega}=\frac{3}{16\pi}\sigma_\mathrm{T}\left(\frac{E_\mathrm{f}}{E_\mathrm{i}}\right)^2
\left( \frac{E_\mathrm{i}}{E_\mathrm{f}} + \frac{E_\mathrm{f}}{E_\mathrm{i}} -\sin^2\theta_s \right),
\eeq 
where 
$E_\mathrm{i}$ is the photon energy before a scattering event,  
\beq\label{eq:Ef_compton}
E_\mathrm{f}=\frac{E_\mathrm{i}}{1+E_\mathrm{i}(1-\cos\theta)/(m_\mathrm{e}c^2)}
\eeq 
is the photon energy after the scattering and $\theta_s$ is the angle between the initial and final photon momentum (see Chapter 7 in \citealt{1986rpa..book.....R}).
Due to each scattering, photons transfer energy
\beq\label{eq:DeltaE}
\Delta E=E_\mathrm{i}-E_\mathrm{f}
\eeq 
and momentum along magnetic field lines 
\beq\label{eq:DeltaP}
\Delta p_{||}=(\bm{p}_\mathrm{i}-\bm{p}_\mathrm{f})\cdot \bm{n}_B, 
\eeq 
where $\bm{p}_\mathrm{i}$ and $\bm{p}_\mathrm{f}$ are photon momenta before and after the scattering event, and 
\beq
\bm{n}_B = 
\left(\begin{array}{c} 
-\sin(\chi+\theta)\cos\varphi \\ 
-\sin(\chi+\theta)\sin\varphi \\
-\cos(\chi+\theta)
\end {array}\right)
\eeq
is the tangent
to the magnetospheric accretion flow.
The transferred energy (\ref{eq:DeltaE}) is going to acceleration/deceleration of accretion flow and its heating/cooling.

In the strong-field regime, the Compton scattering is much more complicated: it becomes resonant and polarization-dependent. 
The general expressions for the Compton scattering matrix in a strong magnetic field have been derived by \citet*{MushtukovNP16}.
In this paper, we adopt a simplified treatment of resonant scattering in the vicinity of a NS, which is described in detail in Appendix~\ref{app:Resonant_scattering}. 
We approximate the scattering cross section using non-relativistic expressions.
This approximation is applied in the inner regions of the magnetosphere, where the magnetic field is strong and resonant effects may potentially influence the energy exchange.
At the same time, we emphasize that this treatment
does not aim to provide a fully self-consistent description of resonant Compton scattering, but rather to assess its possible impact on the thermal structure of the flow.

\subsubsection{Compressional heating}
\label{sec:Compression}

In a steady, magnetically guided flow confined to a thin flux tube, 
the compressional (adiabatic) heating per unit volume is
\begin{equation}
q^{+}_\mathrm{comp} \;=\; - P\,\nabla\!\cdot\!\boldsymbol{v},
\label{eq:qcomp_vol}
\end{equation}
where $P$ is the gas pressure and $\boldsymbol{v}=v(s)\,\boldsymbol{n}_B$ is the velocity 
along the field line (arc–length coordinate $s$; $\boldsymbol{n}_B$ is the unit tangent).

For a one–dimensional flow along a flux tube of cross–section $A(s)$,
\begin{equation}
\nabla\!\cdot\!\boldsymbol{v}
\;=\; \frac{1}{A}\,\frac{\dd\,(A v)}{\dd s}
\;=\; \frac{\dd v}{\dd s}+\frac{v}{A}\frac{\dd A}{\dd s}.
\label{eq:div_tube}
\end{equation}
Using magnetic–flux conservation $A(s)\,B(s)=\text{const}$, one can eliminate $A(s)$:
\beq\label{eq:B_cons}
\frac{1}{A}\frac{\dd A}{\dd s} = -\frac{1}{B}\frac{\dd B}{\dd s}
\eeq 
and thus
\beq 
\nabla \cdot \boldsymbol{v}
= \frac{\dd v}{\dd s}-\frac{v}{B}\frac{\dd B}{\dd s},
\label{eq:div_via_B}
\eeq
which is convenient here since $B(r,\theta)$ is known for the dipole geometry (\ref{eq:B}) and $s$ is related to $(r,\theta)$ by equation~(\ref{eq:dx}). 

Integrating equation~\eqref{eq:qcomp_vol} across the local geometrical thickness $H$ of the flow 
and introducing the surface density $\Sigma=\rho H$, we obtain the heating per unit area:
\beq
\label{eq:Qcomp_area}
Q^{+}_{\mathrm{comp},H}=\int_0^H q^{+}_\mathrm{comp}(s)\,\dd s 
\simeq 
-\,\frac{T\,\Sigma}{ m_\mathrm{p}}\,\left(\frac{\dd v}{\dd s}-\frac{v}{B}\frac{\dd B}{\dd s}\right).
\eeq
In a converging flow ($\dd(Av)/\dd s<0$) the term $Q^{+}_{\mathrm{comp},H}$ is positive, i.e. 
compression heats the gas.
For further use, it is convenient to rewrite equation~(\ref{eq:Qcomp_area}) as follows:
\beq
Q_\mathrm{comp,H}^{+} \simeq  2.9\times
10^{16}\Sigma T_{100}
 \left[ \frac{\beta_0}{B_6}\frac{\dd B_6}{\dd s_8}
-\frac{\dd\beta_0}{\dd s_8} \right]
\; \mathrm{erg~cm^{-2}\,s^{-1}} \!,
\label{eq:Q_compression}
\eeq
where $\beta_0\equiv v/c$ 
is the dimensionless bulk plasma velocity introduced above and $s_8\equiv s/10^8$~cm.

\subsection{Ionization degree of accretion flow}
\label{sec:Ionization_degree}

\subsubsection{LTE approximation}
\label{sec:LTE}

Our estimations of the ionization degree are based on the assumption that {the particles and atoms in the accretion flow obey the Boltzmann distribution, which is justified if} the time scale {needed} to acquire {thermal} equilibrium of particles is much shorter than the time scale of variations of thermodynamic characteristics of the flow along its way towards NS surface.
{This condition may be called a partial local thermodynamic equilibrium (LTE). 
The full LTE, which includes also the Planck distribution of photons, is a sufficient but not necessary condition to validate this assumption.
Let us show that the partial LTE likely holds under typical conditions in the accretion channel of an XRP.}

{First, let us consider the case without a strong magnetic field, which is realized far enough from the NS surface.
In this case, the characteristic} relaxation time scales for {non-degenerate} electrons {and protons} can be estimated {as follows} (\citealt{Spitzer_book}, Chapter~5):
\begin{align}&
t_\text{c,e-e}\sim \frac{m_\mathrm{e}^{1/2}T^{3/2}}{n_\mathrm{e}e^4}
\sim
10^{-10}\,\frac{T^{3/2}_{100}}
{n_\mathrm{e,19}}\,\,\mathrm{s},
\\&
t_\text{c,p-p} \approx \sqrt{\frac{m_\mathrm{p}}{m_\mathrm{e}}} \,t_\text{c,e-e}\sim 
4\times
10^{-9}\,\frac{T^{3/2}_{100}}{n_\mathrm{e,19}}\,\,\mathrm{s},
\label{t_cpp}
\\&
t_\text{eq,e-p} \approx \frac{m_\mathrm{p}}{m_\mathrm{e}} \,t_\text{c,e-e} \sim 
2\times
10^{-7}\,\frac{T^{3/2}_{100}}
{n_\mathrm{e,19}}\,\,\mathrm{s},
\end{align}
where $t_\text{c,e-e}$ ($t_\text{c,p-p}$) is the time required for the electrons (protons) to acquire a Maxwellian distribution, and $t_\text{eq,e-p}$ is the time scale of energy exchange between the electrons and protons, at which their temperatures become equal.

In the case of neutral atoms, a typical collisional cross section is{, by order of magnitude,
$\sigma_\mathrm{neut}\sim 10\,a_\mathrm{B}^2\sim 10^{-15}\,\mathrm{cm^2}$
\citep[cf.][Chapter~1]{1992pavi.book.....S}, where $a_\mathrm{B} = \hbar^2/m_\mathrm{e} e^2 \approx 0.529\times10^{-8}$~cm is the Bohr radius. Therefore} a time scale between collisions is
\beq 
t_\text{neut} \sim \frac{1}{v_{T,\text{p}} n \sigma_\mathrm{neut}} 
\sim
10^{-11}\,T_{100}^{-1/2}
n_{19}^{-1}\,\,\text{s},
\label{t_neut}
\eeq 
where $v_{T,\text{p}}\simeq
10^7\,T_{100}^{1/2}
\,\,\mathrm{cm\,s^{-1}}$ is the thermal velocity of protons and H atoms.

These time scales are orders of magnitude shorter then the free-fall time from the inner edge of an accretion disc {at $r=R_\mathrm{m}$ to the inner radius of the weak-field zone $R_*\sim10 R$ in a typical XRP, hence the Boltzmann distribution of particles is established} for the magnetospheric accretion flow.  

{In the strong-field regime, at the densities and temperatures typical for the XRP accretion channels, spontaneous radiative decay of the excited electron Landau levels is orders of magnitude faster than Coulomb excitations, therefore it is a good approximation to assume that all electrons reside on the ground Landau level (see, e.g., \citealt{Meszaros_book}, Section~4.4).
However, the magnetic quantization affects only the transverse motion, hence
the longitudinal velocity distributions of the particles injected into the strong-field zone remains to be Maxwellian.
An order-of-magnitude estimate of the velocity-averaged transition rate for non-radiative Coulomb collisions between such electrons is \citep[cf.][]{PotekhinLai07}
$
   \Gamma_\mathrm{c,ee} \sim 4\sqrt{\pi}\,n_\mathrm{e} a_\mathrm{m}^3 \,\tilde\Lambda\,\tau_0^{-1}
$,
where  $a_\mathrm{m} = \sqrt{\hbar c/eB} = a_\mathrm{B}\sqrt{B_0/B}$ is the magnetic length,
$\tau_0 = \hbar^3/m_\mathrm{e} e^4 \approx 2.42\times10^{-17}$~s is the atomic unit of time and $\tilde\Lambda$ is a modified Coulomb logarithm, which is of the order of 0.1--10 in a strongly quantizing magnetic field.
Hence a typical electron-electron collision time in the strong-field regime is by order of magnitude
\beq
   t_\text{c,e-e}\sim 10^{-10}\frac{B_{12}^{3/2}}{n_\text{e,19}}\text{~s}.
\label{t_cee_mag}
\eeq
If the magnetic field is weakly quantizing for protons, then the non-magnetic formula (\ref{t_cpp}) can be used for rough estimates. In the case of strongly quantized protons,
$
   t_\text{c,p-p}^\text{mag}\sim (m_\mathrm{p}/2m_\mathrm{e}) \,t_\text{c,e-e}
   \sim 10^{-7}\,(B_{12}^{3/2}/n_\text{e,19})$~s.
}

{The mean squared size of an H atom
in a strong magnetic field is inversely proportional to its binding energy $E_\mathrm{b}$, which increases roughly as $\ln^2(B/B_0)$ (an accurate fit is given in Appendix~\ref{app:PhotoIonCS}).
Therefore $\sigma_\mathrm{neut}$ decreases and $t_\text{neut}$ increases as $\sim\ln^2(B/B_0)$, which is $\lesssim10^2$, if $B \lesssim 10^{14}$~G. Thus on the base of equation~(\ref{t_neut}) we have $t_\text{neut} \lesssim 10^{-9}\,T_{100}^{-1/2}
n_{19}^{-1}\,\,\text{s}$.
}

All these time estimates are much shorter than a typical plasma free-fall time through the strong-field zone
$t_* \sim R_*/v \sim 10^{-3}$~s. Therefore, the partial LTE conditions can be fulfilled in this zone.

When the accreted plasma enters the strong-field regime, the increase of the binding energies can result in recombination of the atoms. The average rate $\Gamma_\mathrm{rec}(n_\mathrm{e},T)$ of spontaneous recombination in thermal equilibrium is estimated in Appendix~\ref{app:rates}.
Since the ambient radiation does not affect this recombination process, the same estimate is valid in the partial LTE without requiring the full equilibrium. Intersections of the solid and dashed lines in Fig.~\ref{pic:rates} occur at the temperatures where the LTE estimate provides the ionization degree $f_+ = 0.5$. At temperatures around or below these values, the characteristic recombination time $\Gamma_\mathrm{rec}^{-1} \lesssim 10^{-6}$~s is several orders of magnitude shorter than $t_*$. Therefore it is a good approximation to assume that, whenever the LTE estimate provides a substantial neutral fraction $f_\mathrm{H}=1-f_+$, recombination to this level occurs instantaneously, as the plasma enters the strong-field regime.

\subsubsection{Ionization degree in a strong magnetic field}
\label{sec:ioniz}

Since the strong magnetic field ($B \gg B_0$)
{increases binding energies of atoms \citep*[e.g.,][]{CohenLR70,KadomtsevKudryavtsev71}},
it also
{suppresses plasma ionization}
(\citealt{Khersonskii87}; see, e.g., \citealt{2014PhyU...57..735P} for review and references). 
Here we evaluate the ionization degree following the method developed by \citet*{PotekhinCS99} in frames of the chemical picture of plasmas, where the basic species are considered to be ions, electrons and atoms (we neglect molecules and molecular ions for simplicity).
The chemical picture of plasmas is tractable easier than the alternative physical picture, in which only the electrons and atomic nuclei are considered as fundamental species, but 
the necessity to distinguish
the bound and free electrons and attribute the bound electrons
to certain nuclei introduces some ambiguity 
(see, e.g., \citealt{Rogers00}, for discussion of strengths and weaknesses of the chemical and physical pictures of plasmas). Current approaches to the solution of this problem are based,
as a rule, on the concept of occupation probabilities of quantum states \citep[e.g.,][]{HummerMihalas88}. 

In the case of strong magnetic fields, the occupation probabilities depend not only on the discrete quantum
numbers, but also on the transverse component $K_\perp$ of the pseudomomentum $\bm{K}$, which is the conserved quantum-mechanical quantity that characterizes the state of motion across an external magnetic field (see, e.g., \citealt*{JohnsonHY83}, for review). The states of
H atoms are conventionally characterized by two discrete quantum numbers, $s$ {(related to the $\bm{B}$-projection of the 
angular momentum)} and $\nu$ {(related to longitudinal wave-function behaviour),} and by $K_\perp$, which has a continuous  distribution $p_{s,\nu}(K_\perp)$. This distribution is not known in advance,
but is obtained as a part of the solution of the ionization equilibrium problem by minimization of the Helmholtz free energy $F$ with respect to particle numbers, keeping constant the total
number density of protons (free and bound) $n_\mathrm{p,tot}$, and the number density of electrons $n_\mathrm{e}$ equal to that of protons because of the overall electrical neutrality. The free energy can be written as
\beq
   F = F_\mathrm{id}^\mathrm{e} + F_\mathrm{id}^\mathrm{p} 
       + F_\mathrm{ex,ep} + F_\mathrm{at} + F_\mathrm{mol} + F_\mathrm{ex,b},
\label{Fren}
\eeq
where $F_\mathrm{id}^\mathrm{e}$,
$F_\mathrm{id}^\mathrm{p}$
are the free energies of ideal gases of the electrons and
protons, respectively, $F_\mathrm{ex,ep}$ arises due to the Coulomb and exchange terms in the fully ionized electron-ion plasma,
$F_\mathrm{at}$ is the contribution of the
atomic gas (including its kinetic and internal degrees of
freedom), $F_\mathrm{mol}$ is the contribution of molecules, and $F_\mathrm{ex,b}$ accounts for interactions of bound species with protons, electrons and with each other. Practical analytical formulae for $F_\mathrm{id}^\mathrm{e}$, $ F_\mathrm{id}^\mathrm{p} $ and $F_\mathrm{ex,ep}$ are collected in \citet{PotekhinChabrier13} and explicit expressions for $F_\mathrm{at}$, $F_\mathrm{mol}$ and $F_\mathrm{ex,b}$ are given in 
\citet{PotekhinCS99}. In particular,
\begin{align}&
 F_\mathrm{at} =
     T V \sum_{s\nu} n_{s\nu}
     \int\limits_0^\infty \ln\left[n_{s\nu} \lambda_\mathrm{H}^3 
        \frac{w_{s\nu}(K_\perp) }{ \exp(1) \mathcal{Z}_{s\nu}}\right]
    \nonumber\\&\qquad\times\,
       p_{s\nu}(K_\perp)2\pi K_\perp\dd K_\perp,
    \label{F-at}
\end{align}
where $V$ is the volume,
$n_{s\nu}$ is the number density of the H atoms with
given discrete numbers $s$ and $\nu$ (any $K_\perp$),
$w_{s\nu}(K_\perp)$ are the occupation probabilities,
$\lambda_\mathrm{H} = [{2\pi\hbar^2}/({ m_\mathrm{H} T})]^{1/2}$
is the thermal wavelength of the H atom, $m_\mathrm{H}=m_\mathrm{e}+m_\mathrm{p}$ is its mass and
\beq
   \mathcal{Z}_{s\nu} =
\frac{ 2\pi }{ m_\mathrm{H} T } 
          \int\limits_0^\infty w_{s\nu}(K_\perp)\,
           \mathrm{e}^{E_{s\nu}(K_\perp)/ T}
            K_\perp \dd K_\perp
\label{Z-int}
\eeq
is the contribution of the states $(s,\nu)$ in the partition function, which {takes  into account the continuous
dependencies of binding energies $E_{s\nu}$ on $K_\perp$}. In equations~(\ref{F-at}) and (\ref{Z-int}) we use the formulae for $w_{s\nu}(K_\perp)$ from \citet{PotekhinCS99}.

It must be noted that interactions between different members of the statistical ensemble in the chemical picture
give rise to a significant fraction of clusters, which contribute to the equation of state 
similarly to the atoms,
but their quantum-mechanical properties (in particular, their cross sections of interactions with radiation) differ from  those of an isolated
atom. Therefore we evaluate the number of bound H atoms excluding the states
that are strongly perturbed by the
plasma environment so that they do not satisfy the  \citet{InglisTeller39I} criterion of
spectral line  merging. Such states form the so called optical
pseudo-continuum  \citep*[e.g.,][]{DappenAM87}. This distinction between the
``thermodynamic'' and ``optical'' neutral fractions is
inevitable in the chemical picture of a plasma 
\citep[see, e.g., the discussion in][]{Potekhin96}. To exclude the strongly perturbed states (the pseudo-continuum),
we apply so called optical occupation
probabilities, which depend on the average atomic size $l_{s,\nu}(K_\perp)$ according to the formula
\beq 
   w_{s\nu}^\mathrm{o}(K_\perp) = \exp\left\{ -(4\pi/3)
      [4 l_{s,\nu}(K_\perp)]^3 n_\mathrm{p} \right\},
\eeq
which is a generalization of equations in section 4 of \citet{HummerMihalas88} to our case of a strong magnetic field.
The fraction of weakly perturbed atoms, which contribute to the
bound-bound and  bound-free opacities, constitutes a
fraction $w_{\nu s}^\mathrm{o}(K_\perp)/w_{\nu s}(K_\perp) < 1$  of
the total number of atoms. Here, $w_{\nu s}(K_\perp)$ is the
``thermodynamic'' occupation probability derived from  the
free energy, which enters the generalized partition function
(\ref{Z-int}).

{Minimization of $F_\mathrm{at}$ leads to the generalized Saha equation (equation 54 in \citealt{PotekhinCS99}). In the case of strongly quantized ($T \ll E_\mathrm{cyc}$), non-degenerate plasma, it reduces to
\beq
n_\mathrm{H} = n_\mathrm{p}n_\mathrm{e} \lambda_\mathrm{e}
\left( \frac{2\pi a_\mathrm{m}^2}{\lambda_\mathrm{H}} \right)^{\!\!2}
\left[ 1 - \ee^{-E_\mathrm{cyc,p}/T} \right]
\mathcal{Z},
\label{magSaha}
\eeq
where $\lambda_\mathrm{e} = \sqrt{2\pi\hbar^2/m_\mathrm{e}T} \approx a_\mathrm{B}\sqrt{1.71/T_{100}}$ is the electron thermal wavelength, $\mathcal{Z} = \sum_{s,\nu}\mathcal{Z}_{s\nu}$ the complete partition function, and we have neglected the difference between $m_\mathrm{p}$ and $m_\mathrm{H}$.
Note that under the charge neutrality condition $n_\mathrm{p}n_\mathrm{e}/n_\mathrm{H} =n_\mathrm{p}^2/n_\mathrm{H} = n_\mathrm{p,tot}f_+^2/(1-f_+)$.
}

In the case where protons are non-quantized ($E_\mathrm{cyc,p}\ll T$), equation~(\ref{magSaha}) reduces to
\beq
n_\mathrm{H} = 2\pi a_\mathrm{m}^2 \lambda_\mathrm{e} n_\mathrm{p}n_\mathrm{e} \mathcal{Z}.
\label{magSaha2}
\eeq
This can be compared with the familiar zero-field Saha equation $n_\mathrm{H} = \frac12 \lambda_\mathrm{e}^3 n_\mathrm{p}n_\mathrm{e} \mathcal{Z}$.

\subsubsection{Photoionization}
\label{sec:Photoionization}

In close proximity to a NS, the ionization degree of the accretion flow can be significantly influenced by interactions with X-ray photons originating from the hot spot on the NS surface via photoionization. 
The photoionization cross section of the ground-state H atom in the weak-field limit (neglecting relativistic and recoil corrections) is
given by the Sommerfeld formula \citep[see, e.g.,][]{Rosmej_20}\footnote{Equations for $\sigma_\mathrm{ph}$ in \citet{Rosmej_20} are valid in atomic units, but in general the factor $c$ in their denominator should be replaced by $\alpha_\mathrm{f}^{-1}$.}
\beq
\sigma_\mathrm{ph}^{(0)}
=
\frac{64\pi}{\alpha_\mathrm{f}^3}\left(\frac{E_0}{E}\right)^{\!\!4} \,
\frac{\exp\big[-4\varsigma\arctan(1/\varsigma)\big]}{1-\exp(-2\pi\varsigma)}
\,\sigma_\mathrm{T},
\eeq 
where $\varsigma = (m_\mathrm{e}e^2/\hbar)\left[2m_\mathrm{e}(E-E_0)\right]^{-1/2}$ is the Born parameter and 
the photon energy $E$ must exceed the ionization threshold
$E_0\approx 1$~Ry, where
$\text{Ry} \equiv \mbox{Ha}/2 = 13.605$~eV is the Rydberg energy.
At the threshold ($E-E_0\to+0$, $\varsigma\to+\infty$), $\sigma_\mathrm{ph}^{(0)}\approx 0.97\times10^7\sigma_\mathrm{T}$.
In the Born limit ($E\to+\infty$, $\varsigma\simeq\sqrt{E_0/E}$\,), $\sigma_\mathrm{ph}^{(0)} \simeq 8.2\times10^7 (E_0/E)^{7/2}\, \sigma_\mathrm{T}$.

In the presence of a strong magnetic field, the photoionization cross-section depends on the direction of photon propagation relative to the local magnetic field and the photon polarization state. 
Unlike the calculations of the ionization degree described in Section~\ref{sec:ioniz}, 
where  the effects of atomic motion across the field have been accurately taken into account, here for simplicity of the estimates we adopt the approximate analytic expressions of \cite{1993ApJ...407..330P}, which account for the effects of a strong magnetic field (see Appendix~\ref{app:PhotoIonCS} for details), but neglect the effects of thermal motion on the photoionization cross-sections (for a treatment of the latter effects, see \citealt{1997ApJ...483..414P}).

The Planck formula gives the average energy per photon in the blackbody radiation equal to $\pi^4 T/30\zeta_3 \approx 2.7 T$, where $\zeta_3$ is the Ap\'ery's constant.
The kinetic energy of a proton near the NS surface is  
$\sim 0.2 m_\mathrm{p}c^2 \approx 
2\times 10^5$~keV.
Thus the number of emitted photons per proton can be roughly estimated from the energy balance as 
\beq
\frac{N_\mathrm{ph}}{N_\mathrm{p}} \sim 
\frac{2\times 10^5\mbox{~keV}}{2.7\, T} \sim
\frac{10^6}{T_{100}}.
\label{ph_to_p}
\eeq

Note that some fraction of the energy released in the NS atmosphere is emitted in the form of cyclotron photons.
These photons undergo further Comptonization, contributing to the high-energy component of X-ray spectra observed at low mass accretion rates. 
Observational evidence for these features has been reported by 
\citet{2019MNRAS.483L.144T,2019MNRAS.487L..30T} and \citet{2021ApJ...912...17L}, with theoretical interpretations provided by \citet{2021MNRAS.503.5193M} and \citet{2021A&A...651A..12S}.

\section{Numerical model}
\label{sec:NumMod}

\subsection{Heating and cooling at the magnetospheric surface}

Because we focus on sub-critical XRPs, the dynamics of magnetospheric accretion flow is weakly affected by the radiation pressure and can be calculated separately from heating/cooling of accretion flow. 
Calculating dynamical structure of the flow (i.e., distributions of the surface density and velocity along magnetic field lines), we follow \citet{2024MNRAS.530..730M}.

On the basis of pre-calculated dynamical structure of the flow,
we start our calculations of the accretion flow temperature.
The temperature structure is calculated in the iterative manner. 
At the beginning, we assume that the temperature of magnetospheric accretion flow is constant all over the magnetosphere of a NS and equal to the effective temperature at the inner disc radius:
\beq \label{eq:T_eff}
T_\mathrm{eff}(R_\mathrm{m})\sim 
30\,B_\mathrm{p12}^{-3/7}L_{37}^{13/28}\,\mathrm{eV}.
\eeq 

Using the assumed temperature distribution, we calculate accretion flow heating per square centimeter of the flow due to the Compton scattering, which we denote $Q^+_\mathrm{sc}H$.
To estimate the heating, we perform a Monte Carlo simulation composed of the following steps: 
\begin{enumerate}[leftmargin=15pt]
\item 
We assume a specific initial distribution of X-ray photons over the directions (i.e., the beam pattern) and energy.
\item\label{step:new_photon}
We simulate three random numbers to determine initial photon energy and direction of motion. 
\item 
Assuming the Schwarzschild metric, we trace each photon to dipole magnetospheric surface and find the coordinates where the photon reaches the magnetosphere and the angle $\theta_B$ between photon momentum and local direction of
the magnetic field
lines.
\item 
Because the dynamical structure of accretion flow is already calculated, we know local surface density of the flow and its optical thickness across magnetic field lines $\tau_\mathrm{e}$. 
Thus, we get the probability of photon scattering by the electrons in the flow, which equals $1 - \exp(-\tau_e/\sin\theta_B)$.
Generating a random number, we decide depending on its value, whether the photon passes through the flow or is scattered in it. 
If the photon passes through the flow, we return to step \ref{step:new_photon}.
If the photon is scattered by the flow, we move on to step \ref{step:ph_redirection}. 
\item\label{step:ph_redirection} 
We choose randomly an electron from the distribution (\ref{eq:Maxwell_moving_electrons}), and on the base of two extra random numbers we determine the direction of the scattered photon and its energy (\ref{eq:Ef_compton}).
Using equations (\ref{eq:DeltaE}) and (\ref{eq:DeltaP}), we calculate the momentum and energy exchange.
We specifically account for the momentum exchange along magnetic field lines because it results in a fractional transfer of photon energy into kinetic energy of accretion flow and slightly reduces the energy that goes into heating of the magnetospheric flow. 
\end{enumerate}
Each run of the simulation includes $10^7$ photons and provides a map of $Q^+_\mathrm{sc}H$.

Using the calculated maps of \(Q^{+}_{\rm sc}H\), \(Q^{+}_{\rm comp}H\), and the radiative cooling terms, we recalculate the temperature structure of the flow. 
In the fiducial calculation we retain the approximate cyclotron cooling term described in Section~\ref{sec:free-free_emission}, while Appendix~\ref{sec:Magnetic_free-free_cooling_new} presents a limiting test in which this term is removed. 
The procedure is iterated until convergence to a stable temperature distribution.

\subsection{Influence of photoionization}

In order to account for photoionization and determine the ionization degree of the accretion flow near the NS surface, we simulate radiative transfer through the accretion channel close to its base. 
We assume the base of the accretion channel is a ring with a geometrical thickness \(d\) and $R_\mathrm{b}\gg d$ accretion channel base radius. 
The radiative transfer simulation is confined to a region with height \(h_\mathrm{max} = 10d\), beyond which photoionization effects become insignificant due to dilution.
We further assume that atoms photoionized near the NS surface do not undergo recombination before reaching the stellar surface.
Consequently, in the region of height $h_\mathrm{max}$ near the surface, the ionization degree either increases or remains unchanged.

The ionization degree is expected to vary with height above the surface and is computed iteratively. 
Each iteration step is indexed by a number $i$. 
In our numerical model, the base region of the accretion channel, up to the height $h_\mathrm{max}$, is divided into $N_\mathrm{l}=500$ layers of equal geometrical thickness \(\Delta h = {h_\mathrm{max}}/{N_\mathrm{l}}\). 
At each iteration step, we estimate the ionization degree \(I^{(i)}(h_j) \in [0,1]\) for the \(j\)-th layer, where \(h_j = (j-1/2)\,\Delta h\) is the height above the NS surface. 
Since we neglect hydrogen recombination, we enforce the condition \(I^{(i)}(h_j) \geq I^{(i)}(h_{j+1})\).

As an initial guess, the ionization degree is assumed to be constant and equal to the value corresponding to the flow temperature near the NS surface. 
For a given ionization degree profile \(I^{(i)}(h)\) and mass accretion rate, we perform a Monte Carlo simulation of radiative transfer through the lower part of the accretion channel, taking into account photon absorption via photoionization. 
Each simulation run involves \(N = 10^7\) photons and proceeds through the following steps:
\begin{enumerate}[leftmargin=15pt]
\item 
Using equation (\ref{eq:T_eff_sp}), we estimate the effective temperature of the hot spots for a given accretion luminosity and surface magnetic field strength. Assuming that the hot spots emit blackbody radiation, we then construct a cumulative photon energy distribution.
\item \label{step:new_ph_PI}
We generate four random numbers $X_i\in(0,1)$, and using the cumulative distribution of photon energies, we obtain the photon energy $E$ in the NS surface reference frame, the specific emission location at the base of the accretion channel, and the photon momentum direction.
\item 
Next, we account for the Doppler effect and relativistic aberration to transform the photon energy and the angle between the photon momentum and magnetic field lines (\(\theta'_B\)) into the reference frame co-moving with the flow:
\beq
E' = E\gamma_0(1+\beta_0\cos\theta_B),\quad
\cos\theta'_B = \frac{\cos\theta_B-\beta_0}{1-\beta_0\cos\theta_B}.
\eeq
\item 
We then calculate the photoionization cross-section in the flow’s co-moving frame and determine the optical depth \(\tau_\mathrm{PI}\) of the accretion channel along the photon trajectory:
\beq
\tau_\mathrm{PI} \simeq \sum\limits_{j=1}^{N_{l,\mathrm{max}}}\left[1-I^{(i)}(h_j)\right]
\frac{n_\mathrm{p}\sigma_\mathrm{PI}(E',\theta'_B)\Delta h}{\cos\theta_B}.
\eeq
Given that \(h_\mathrm{max}\ll R\), we neglect general relativistic effects such as gravitational redshift and bending.
\item 
By generating a random number \(X_i\), we obtain the photon’s free path optical depth: 
\(\tau_\mathrm{fp}=-\ln X_i\).
If \(\tau_\mathrm{fp} > \tau_\mathrm{PI}\), the photon escapes the accretion channel without interaction with matter, and we return to step \ref{step:new_ph_PI}.
Otherwise, the photon is absorbed through photoionization, and we proceed to step \ref{step:absorption}.
\item
\label{step:absorption}
Using the optical depth contributions from all layers along the photon’s path and the photon’s free path, we identify the specific layer where photon absorption occurs. After accounting for photon absorption in this layer, we return to step \ref{step:new_ph_PI}.
\end{enumerate}
Upon completing the simulation, we determine the fraction of photons absorbed in each layer, denoted as \(f_{\mathrm{abs},j} = N_{\mathrm{abs},j}/N\), where \(N_{\mathrm{abs},j}\) represents the number of photons absorbed in the \(j\)-th layer out of \(N\) simulated photons. Using these absorption fractions, we then correct the ionization degrees for each layer from the current simulation step.
The ionization degree \(I^{(i)}(h_j)\) in the \(j\)-th layer is increased if (see equation \ref{ph_to_p})
\[
I^{(i)}(h_j) - I^{(i)}(h_{j+1}) < \frac{N_\mathrm{ph}}{N_\mathrm{p}}f_{\mathrm{abs},j}
\]
and decreased otherwise. 
We continue iterating until the ionization degree increments match the fraction of absorbed photons in each layer, yielding the following condition for convergence:
\beq 
I(h_j) - I(h_{j+1}) \simeq \frac{N_\mathrm{ph}}{N_\mathrm{p}}f_{\mathrm{abs},j}.
\eeq

\smallskip\smallskip\smallskip
The numerical calculations are performed using custom codes developed by the authors. 
The implementation has been verified through a series of consistency checks, including comparison with analytical estimates in limiting regimes (e.g. optically thin cooling and Compton heating) and by reproducing the expected scaling behaviour of the temperature and ionization structure.

\begin{figure}
\centering
\includegraphics[width=\columnwidth]{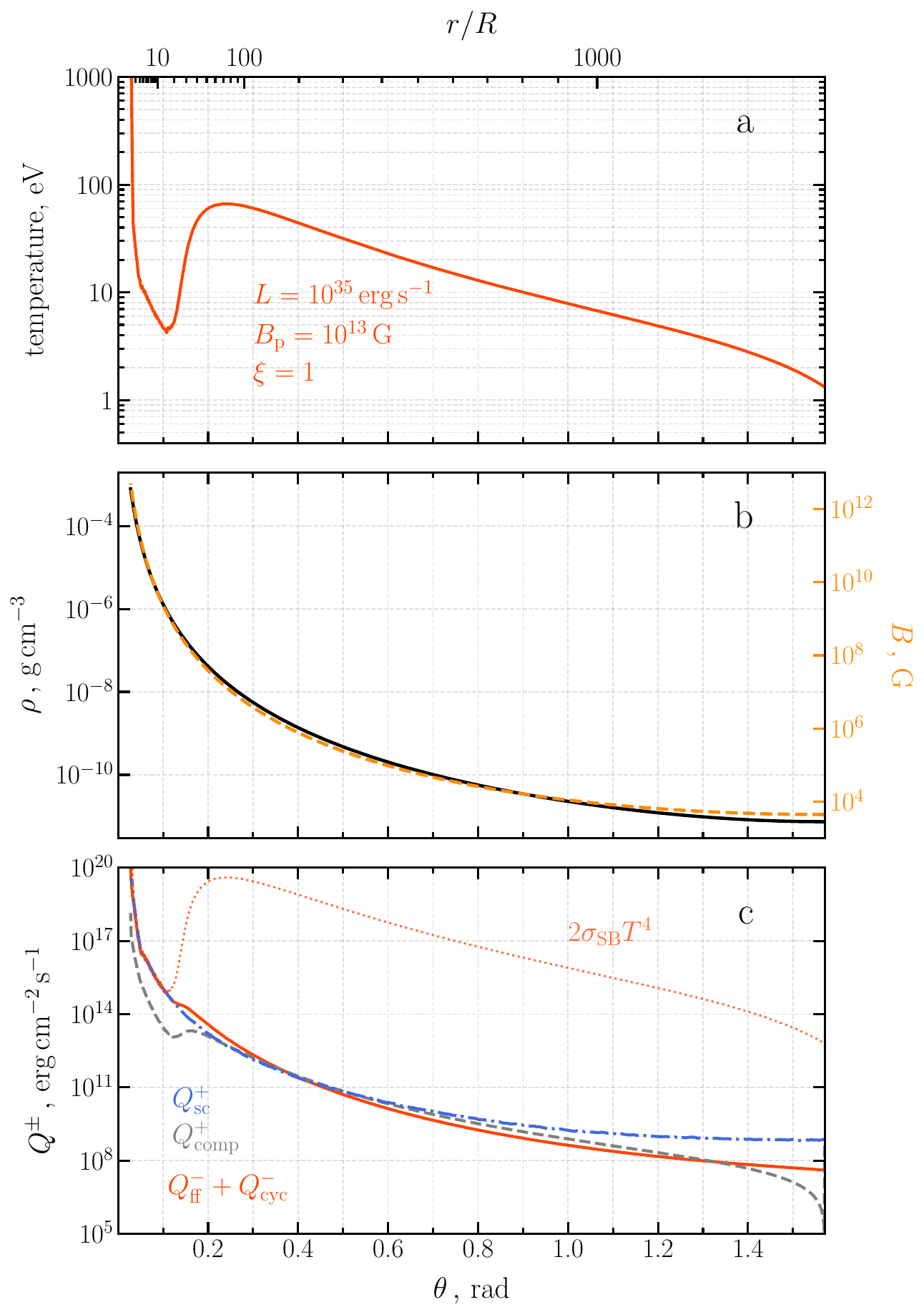}
\caption{
Profiles of the thermal balance in the magnetospheric accretion flow.
Panel a: temperature of the flow as a function of the coordinate $\theta$ along the magnetic field line. For the reference, the upper horizontal axis shows the radial coordinate in units of NS radii. 
Panel b: the mass density (solid black line) and local $B$-field strength (dashed red line) in accretion channel.
Panel c: heating and cooling rates as functions of the same coordinate $\theta$.
The blue dash--dotted line shows Compton heating by X-ray photons from the NS surface ($Q^+_\mathrm{sc}$), 
the grey dashed line corresponds to compressional heating ($Q^+_\mathrm{comp}$), 
and the red solid line represents radiative cooling due to free--free and cyclotron emission ($Q^{-}_\mathrm{ff}+Q^{-}_\mathrm{cyc}$).
The thin dotted line shows the blackbody cooling rate, $2\sigma_{\rm SB}T^4$, corresponding to emission from both sides of a plane-parallel layer at the local flow temperature.
Where the radiative cooling curve approaches this line, the losses in the fiducial calculation are limited by the imposed blackbody flux.
}
\label{pic:sc_Q}
\end{figure}

\begin{figure}
\centering 
\includegraphics[width=\columnwidth]{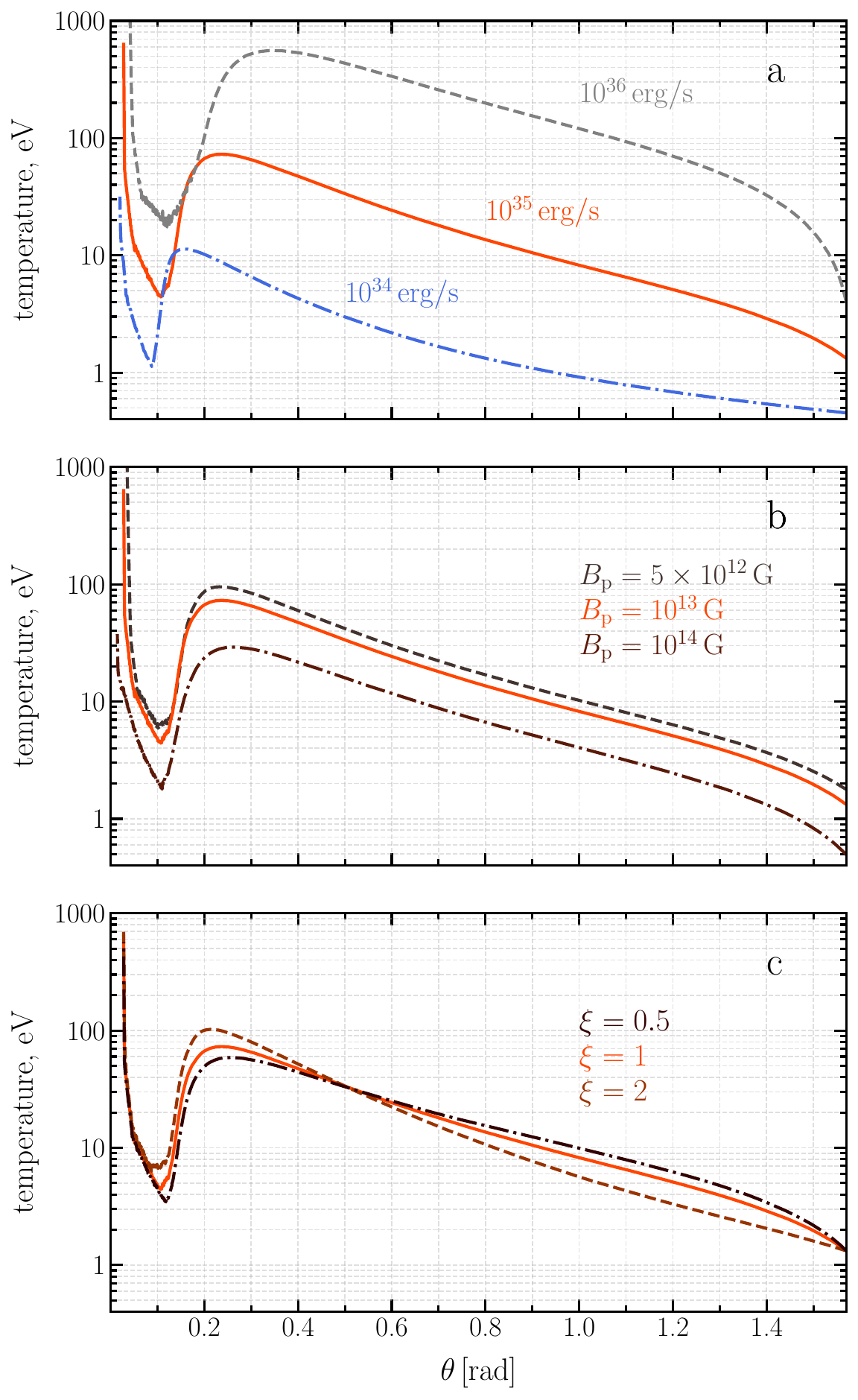}
\caption{
Temperature distribution along the accretion flux trajectory in the magnetosphere of an XRP, parametrized by the magnetic colatitude $\theta$ (see Fig.~\ref{pic:dipole_scheme}A).
The fiducial case is given by solid red line and corresponds to the accretion luminosity
$L=10^{35}$ \ergs, magnetic field at the pole $B_\mathrm{p}=10^{13}\,\text{G}$ and beaming parameter in equation~\eqref{eq:beam} $x$.
NS mass and radius are fixed at $M=1.4M_\odot$ and $R=10^6\,\mathrm{cm}$.
Panel (a) illustrates the influence of accretion luminosity. Dashed grey and dashed-dotted blue lines are calculated for $L=10^{36}$ \ergs\ and $10^{34}$ \ergs\ respectively.
The lower the luminosity, the lower the temperature of material covering the NS magnetosphere.
Panel (b) demonstrates the influence of NS magnetic field strength, which affects the size of the magnetosphere. Dashed and dot-dashed lines here are given for $B_\mathrm{p}=5\times 10^{12}$~G and $10^{14}$~G respectively.
Increasing magnetic field enlarges the magnetosphere, reducing its optical thickness and decreasing the efficiency of heating by Compton scattering.
Panel (c) illustrates the influence of X-ray beam pattern, which naturally affects the distribution of heating due to the scattering.
}
\label{pic:sc_flow_temp_}
\end{figure}

\section{Numerical results}
\label{sec:NumRes}

The major parameters of numerical simulations are accretion luminosity, beam pattern of the emitted photons, their energy distribution, size of the magnetosphere.

\subsection{Temperature distribution}

The temperature of the magnetospheric accretion flow is determined by the balance between heating due to Compton scattering and compressional heating, and cooling due to free–free and cyclotron emission.
An example of the resulting thermal structure is shown in Fig.~\ref{pic:sc_Q}. 
The top panel presents the temperature profile along the magnetic field lines, while the bottom panel shows the corresponding heating and cooling rates. 
One can see that the absolute values of both heating and cooling rates tend to increase towards the NS as the flow approaches the stellar surface.
In the regions located close to the NS, the thermal balance is primarily determined by the competition between Compton heating and radiative cooling dominated by free–free emission.

The treatment of cyclotron cooling in the strongly quantizing region is uncertain because a direct LTE prescription can overestimate the true photon production rate. 
We examine this uncertainty in Appendix~\ref{sec:Magnetic_free-free_cooling_new} by repeating the calculation with the cyclotron cooling term removed. 
The resulting near-surface temperature is almost unchanged, confirming that the temperature minimum relevant for recombination is mainly produced by the increase of density and the associated enhancement of free-free cooling.

At accretion luminosities $L \gtrsim 10^{35}$ \ergs, heating of the accretion flow due to Compton scattering dominates over the cooling. 
As a result, the temperature of the flow increases as material moves toward the NS along field lines (see Fig.~\ref{pic:sc_flow_temp_}).
In close proximity to the NS, however, the temperature of the flow decreases first due to the sharp rise in particle number density, which enhances the efficiency of cooling processes due to free-free emission ($Q_\mathrm{ff}^- H\propto \Sigma^2$ according to equation \ref{eq:Q_min_1}). 
Then the temperature increases again because the flow becomes optically thicker and scatters a larger fraction of X-ray photons. 
The temperature profile is affected by the mass accretion rate  (Fig.\,\ref{pic:sc_flow_temp_}a), 
the magnetic field strength of NS (Fig.\,\ref{pic:sc_flow_temp_}b), 
and the beam pattern of the X-rays emitted on the NS surface  (Fig.\,\ref{pic:sc_flow_temp_}c):
\begin{itemize}[leftmargin=15pt]
\item[(a)]
At a larger accretion luminosity, the flow intercepts a larger fraction of energy due to the Compton scattering, which effectively heats the flow up. 
At $L\sim 10^{35}$ \ergs, the temperature of the flow near the NS surface is expected to reach $\sim 100\,\mathrm{eV}$. 
At a lower luminosity of $L\sim 10^{34}$ \ergs, the flow temperature above the polar caps may be as low as a few eV.
\item[(b)]
For a stronger magnetic field, which corresponds to a larger magnetosphere, the heating becomes less efficient due to a reduction of the optical thickness of the flow.
As a result, the stronger the field of a NS, the lower the temperature of the flow (Fig.\,\ref{pic:sc_flow_temp_}b).  
\item[(c)]
The beam pattern determines the spatial distribution of Compton heating across the NS magnetosphere. 
In the case of a fan beam pattern, where radiation is suppressed along the normal to the NS surface (corresponding to a smaller parameter $\xi$ in Fig.\,\ref{pic:sc_flow_temp_}c), more efficient heating is expected closer to the plane of the accretion disc.  
\end{itemize}  

We note that in the calculations above we assumed a blackbody spectrum of the surface emission. 
Observed spectra of XRPs are typically harder and are often described by a power law with an exponential cutoff. 
To test the sensitivity of our results to the assumed spectral shape, we repeated the calculations using such spectra, normalized to the same bolometric luminosity. 
We find that the resulting temperature profiles differ only moderately, with the temperature near the neutron star surface increasing by at most a factor of $\sim 2$ (see Fig.~\ref{pic:sc_flow_temp_v6}b). 
The details are presented in Appendix~\ref{app:Spectra}.

\begin{figure}
\centering 
\includegraphics[width=\columnwidth]{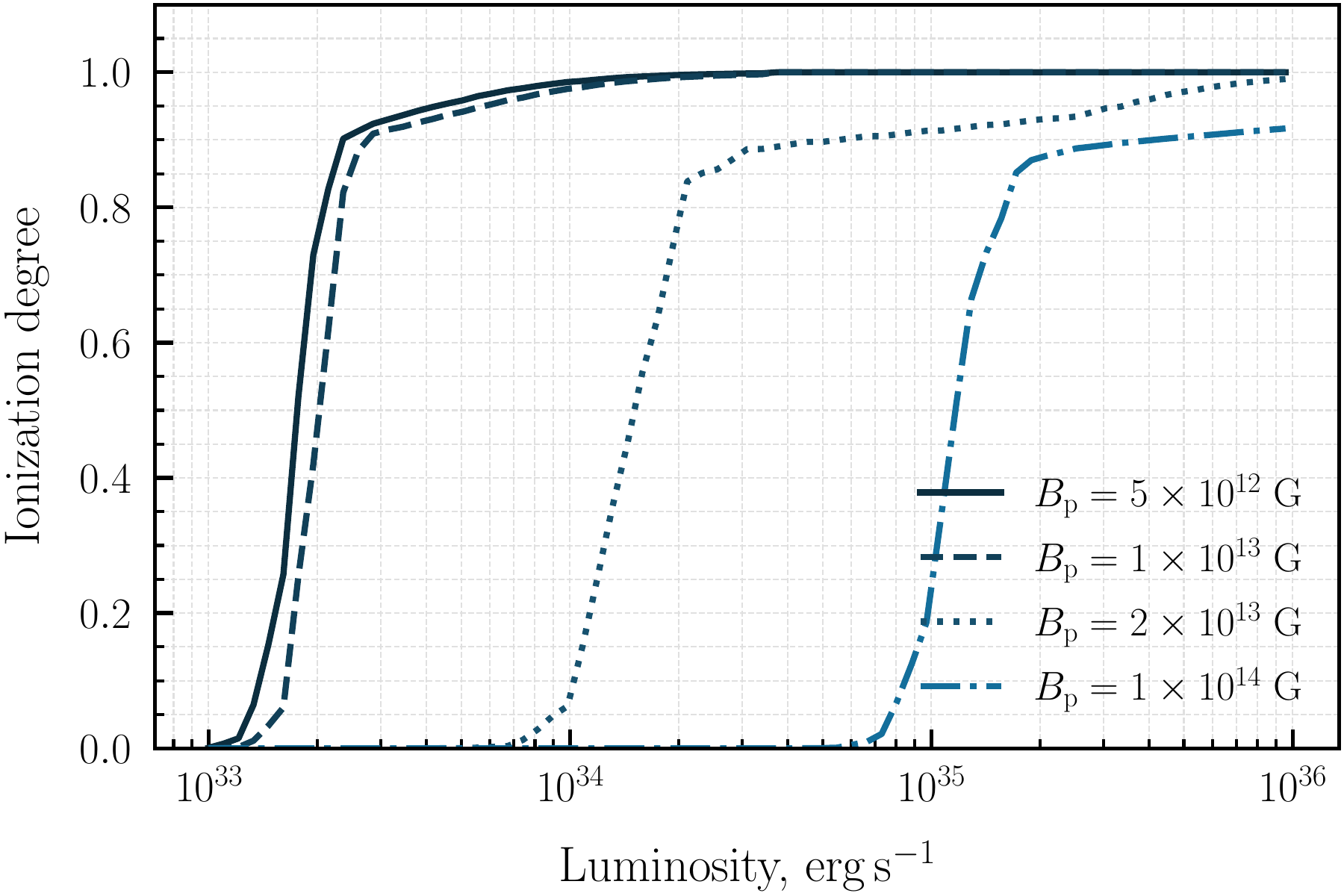}
\caption{
Ionization degree of the accretion flow above the NS surface (at height $h\sim 3d$), computed assuming the Boltzmann distribution, as a function of accretion luminosity. 
Curves correspond to different surface magnetic fields: $B_\mathrm{p}=5\times10^{12}$\,G (solid), $1\times10^{13}$\,G (dashed), $2\times10^{13}$\,G (dotted), and $1\times10^{14}$\,G (dash--dot). 
At sufficiently low luminosities the flow remains weakly ionized. 
Above a characteristic transition luminosity it becomes highly ionized. 
The transition luminosity increases with the magnetic field. 
}
\label{pic:sc_IO}
\end{figure}

\subsection{Ionization degree above the NS surface}
\label{sec:Ionisation_degree_above_NS}

At accretion luminosities below $10^{35}$ \ergs\ ($10^{34}$ \ergs), the temperature of the material in the accretion channel above the NS surface drops below $100\,\mathrm{eV}$ ($10\,\mathrm{eV}$), which is already sufficient to trigger hydrogen recombination in the presence of a strong magnetic field. 
Since the atomic binding energy increases with magnetic field strength, {recombination occurs below a characteristic luminosity that depends on the field strength}
(see Fig.\,\ref{pic:sc_IO}). 
For $B_\mathrm{p}\sim 5\times 10^{12}$~G, recombination occurs at $L\lesssim 2\times 10^{33}$ \ergs, whereas for $B_\mathrm{p}\sim 2\times 10^{13}$~G this characteristic luminosity increases to $L\sim 10^{34}$ \ergs.

Fig.~\ref{pic:sc_flow_temp_v8} shows the ionization structure of the flow over intermediate radii, \(r\lesssim 20\)--\(25R\), for several accretion luminosities at fixed \(B_{\rm p}=10^{13}\,{\rm G}\).
For each point along the flow, the ionization degree is calculated using the local temperature and mass density obtained from the thermal-balance calculation.
The figure demonstrates that partial recombination is not restricted only to an infinitesimally thin layer near the stellar surface.
Instead, for low accretion luminosities, a radially extended weakly ionized region appears in the inner magnetosphere.
This happens because the local magnetic field remains sufficiently strong to increase the hydrogen binding energy, while the flow temperature is low enough for a substantial neutral fraction to be maintained.
As the luminosity increases, Compton heating becomes more efficient and the weakly ionized region shrinks.

\begin{figure}
\centering 
\includegraphics[width=\columnwidth]{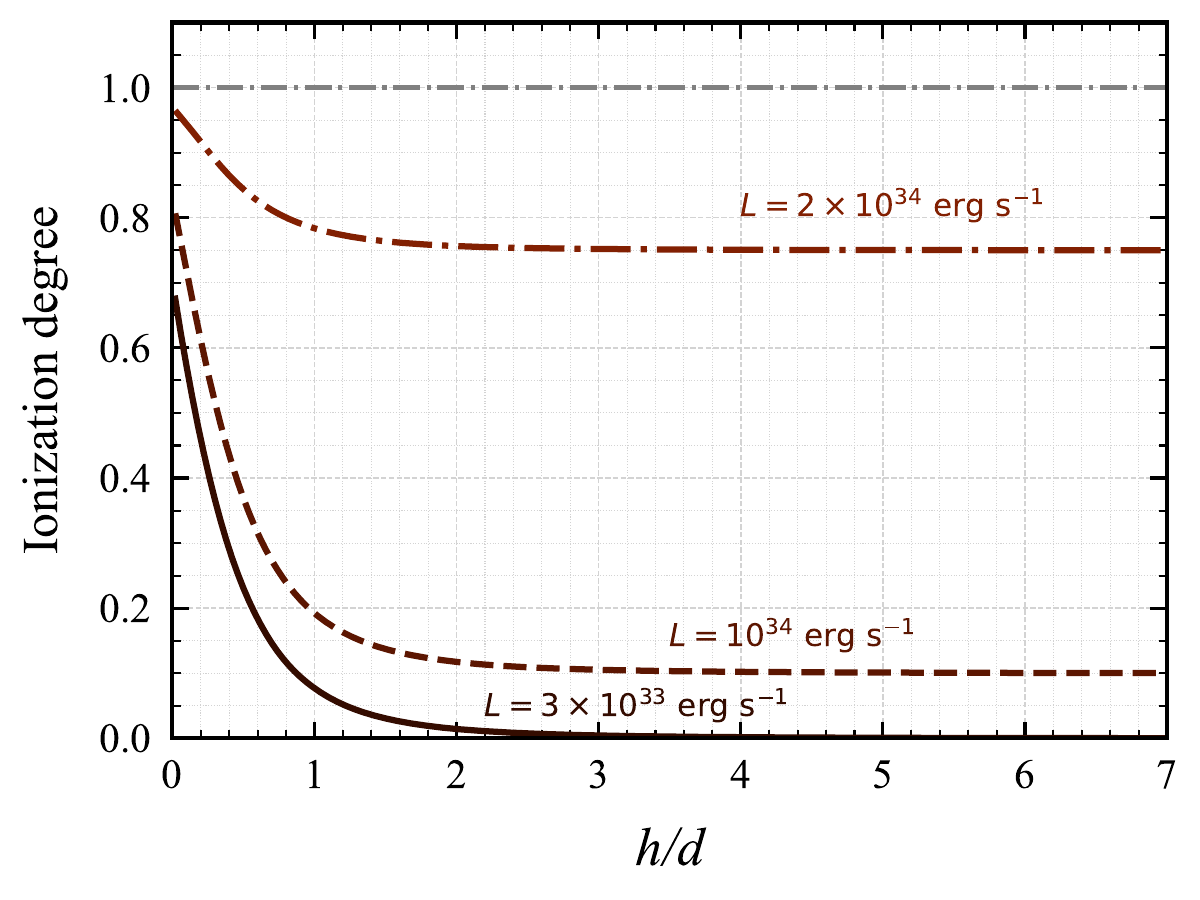}
\caption{
Ionization degree as a function of relative height $h/d$ above the NS surface (see Fig.~\ref{pic:dipole_scheme}). 
The height is given in units of the accretion channel geometrical thickness $d$.
Different curves are calculated for different accretion luminosity: 
$L = 3\times 10^{33}$ \ergs (solid line),
$10^{34}$ \ergs (dashed line),
and 
$2\times 10^{34}$ \ergs (dashed-dotted line).
Surface magnetic field is fixed at $2\times 10^{13}$~G.
Dashed-dotted line at the level of unity corresponds to the case of complete ionization.
At any $L$, there appears a layer with a relatively high ionization degree. 
A thickness of this layer depends on the mass accretion rate.
}
\label{pic:sc_Ph_Ionization}
\end{figure}

\begin{figure}
\centering 
\includegraphics[width=\columnwidth]{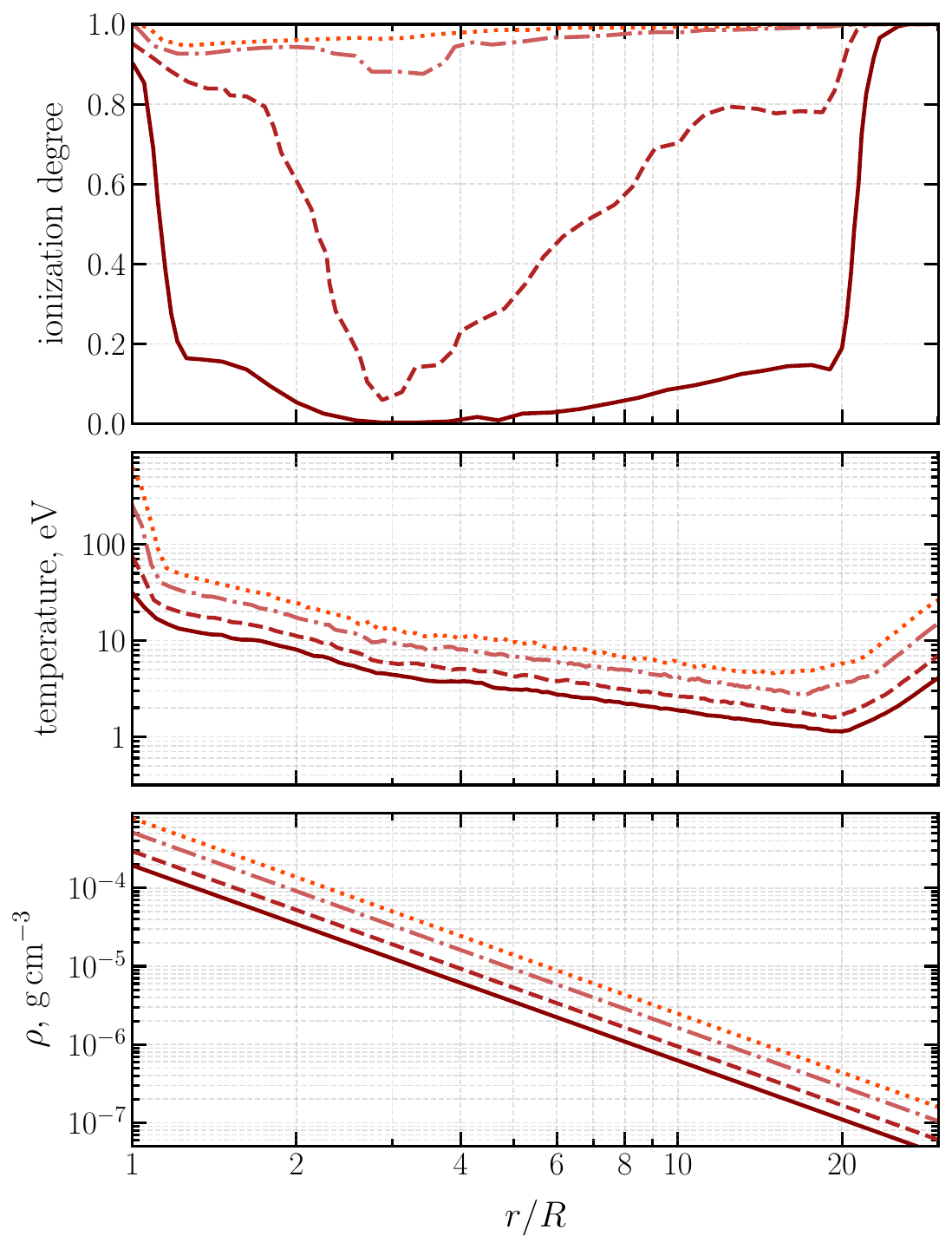}
\caption{
Ionization structure of the flow at intermediate distances from the NS.
The calculations are shown for a fixed surface magnetic field
\(B_{\rm p}=10^{13}\,{\rm G}\) 
and four accretion luminosities:
\(L=10^{34}\,{\rm erg\,s^{-1}}\) (solid line),
\(2\times10^{34}\,{\rm erg\,s^{-1}}\) (dashed),
\(5\times10^{34}\,{\rm erg\,s^{-1}}\) (dashed-dotted), 
and
\(10^{35}\,{\rm erg\,s^{-1}}\) (dotted).
The upper panel shows the ionization degree calculated from the local temperature and density along the magnetospheric flow.
The middle and lower panels show the corresponding temperature and mass density profiles, respectively.
The radial coordinate is given in units of the NS radius.
A broad partially ionized region appears at radii of several to few tens of NS radii, where the local magnetic field is strong enough, for the given temperature, to enhance the hydrogen binding energy and suppress ionization.
}
\label{pic:sc_flow_temp_v8}
\end{figure}

However, an accretion flow in the low-ionization state is subject to photoionization just above the NS surface, which results in formation of a geometrically thin layer in the accretion channel where the ionization degree $f_+$ is higher than at larger distances from the surface (see Fig.\,\ref{pic:sc_Ph_Ionization}). 
With increasing the height $h$ within this layer, $f_+$ drops from a higher value to the nearly constant level presented in Fig.\,\ref{pic:sc_IO}.
The geometrical thickness of the ionized layer near the surface is comparable to the size of the optically thick part of the accretion channel and depends on the accretion luminosity: the higher the luminosity, the larger the thickness of the ionized layer.

Although the ionization level above the re-ionization layer is nearly constant at the scale of Fig.~\ref{pic:sc_Ph_Ionization}, it will increase at large heights $h \gtrsim R$, because the magnetic field weakens with $h$ as $B \sim B_\mathrm{p} /(1+h/R)^3$. For example, if $B_\mathrm{p} \sim 10^{13}$~G, then $B$ falls below the atomic unit $B_0$ at $h\gtrsim 15 R$. In this region, the non-magnetic estimates for the ionization degree apply, giving almost full ionization of hydrogen at any $\rho$, if $T$ exceeds several eV (see, e.g., Fig.~1 in \citealt{Potekhin96}).

{We note that the presence of a partially recombined layer immediately above the NS surface may have implications for the plasma deceleration in the upper atmosphere. 
In particular, a reduced number of free electrons can modify the efficiency of radiative braking and alter the structure of the deceleration region. 
A quantitative assessment of this effect, however, requires a self-consistent treatment of radiative transfer and plasma dynamics in the atmosphere and is beyond the scope of the present work.
At the same time, such conditions are expected to influence the formation of cyclotron features in the accretion channel, since the resonant scattering occurs in the same region above the surface. 
This provides an additional motivation for considering partial recombination effects when interpreting observational data.
}

\section{Summary and discussion}
\label{sec:Summary}

We have performed numerical simulation of cooling and heating of magnetospheric accretion flow in sub-critical X-ray pulsars, where radiative force does not affect dynamics of an accretion flow (i.e., accretion luminosity is $L\lesssim 10^{37}\,\ergs$). 
To estimate the temperature of the flow, we have taken into account free-free cooling, approximate cyclotron cooling, Compton scattering, and compressional heating.
Combining calculations of the dynamics of the accretion flow (performed under the assumption of dipole magnetic field structure; see, e.g., \citealt{2024MNRAS.530..730M}) with its heating and cooling, we have calculated the temperature distribution over the NS magnetosphere and the ionization state of accretion flow reaching the surface of a NS.
The final temperature of the flow is shown to
depend
on accretion luminosity $L$, magnetic field strength and beam pattern of X-ray emission (Fig.~\ref{pic:sc_flow_temp_}).

At $L\lesssim 10^{34}$ \ergs, the temperature of accretion flow above the NS surface drops below a few tens of eV.
Meanwhile, in the strong magnetic fields in the vicinity of the NS surface (typically, $B\sim 10^{12}-10^{13}$~G for the XRPs), the binding energies of the H atoms increase by an order of magnitude.
This results in recombination of a significant fraction of electrons with protons (Section~\ref{sec:ioniz}).

However, accretion flow of a small degree of ionization is still under the influence of photoionization by hard X-ray photons emitted from the NS surface (Section \,\ref{sec:Photoionization}).
The photoionization causes the appearance of a re-ionized layer in the accretion channel (Fig.\,\ref{pic:sc_Ph_Ionization}), where the ionization degree decreases from a higher value near the NS surface to the level determined by the thermal equilibrium of the accretion flow. 
The thickness of this layer is close to the thickness $d$ of the accretion channel across magnetic field lines and depends on $L$: the higher $L$, the higher the thickness of the re-ionized layer.

The ionization state of the accretion flow just above the NS surface can affect the process of spectral formation at low mass accretion rates. 
In particular, the presence and abundance of free electrons above the surface can affect cyclotron scattering features in the spectra of XRPs in low-luminosity states.

The formation of the cyclotron scattering features in the spectra of sub-critical XRPs
can be affected by resonant scattering of X-ray photons by electrons above the NS surface.
Because of the high velocity of the accretion flow above the surface, the scattering feature is expected to be redshifted due to the Doppler effect
(see, e.g., \citealt{2015MNRAS.454.2714M}).
Variations in the velocity of the accretion flow with luminosity in sub-critical XRPs can explain the observed positive correlation between cyclotron line centroid energy and X-ray flux
\citep{2007A&A...465L..25S,2012A&A...542L..28K}.
Recently, it was discovered that in some XRPs, GRO~J1008-57
\citep{2021ApJ...919...33C} and 1A~0535+262
\citep{2024MNRAS.528.7320S}, at very low mass accretion rates, the positive correlation transitions into a negative correlation between the line centroid energy and luminosity.
We note, however, that the luminosities at which partial recombination becomes important in our calculations are lower than those inferred for the observed changes of cyclotron-line behaviour in these sources.
We therefore do not attempt to explain these observations directly.
Instead, we point out that a changing ionization state of the accretion flow provides an additional physical ingredient that may affect the formation of cyclotron features at sufficiently low luminosities.

In addition, partial recombination of the accretion flow may affect the coupling between plasma and the magnetic field in the magnetosphere of a NS. 
In a partially ionized plasma, the magnetic field is primarily coupled to the charged component, while neutral atoms do not directly experience the Lorentz force and interact with the magnetic field
via collisions with charged particles
and via the dependence of the binding energies on the state of atomic motion across the field. The latter effect manifests in the anisotropic kinematics of the atoms (quantified by so called transverse mass; see \citealt{KadomtsevKudryavtsev72,PavlovMeszaros93}): their continuous motion across the field lines slows down, but remains possible.
As a result, the coupling between matter and magnetic field becomes incomplete, and the neutral component can drift relative to the ionized one (ambipolar diffusion; see, e.g., \citealt{2018SSRv..214...58B}). 
This effect may reduce the efficiency of magnetic confinement and potentially modify the geometry and dynamics of the accretion flow close to the stellar surface. 
A quantitative treatment of this effect, however, requires a self-consistent
treatment of the dynamics
of partially ionized magnetized plasma and is beyond the scope of the present work.

The presence of a partially ionized layer may also affect the dielectric properties of the plasma in the accretion channel \citep{2004ApJ...612.1034P}. 
In strong magnetic fields, the polarization properties of radiation are determined by the dielectric tensor of the medium, which depends on the composition and ionization state of the plasma. 
Therefore, even a geometrically thin partially recombined layer may influence the observed X-ray polarization signal. 
A detailed analysis of these effects is left for future studies.

\section*{Acknowledgements}

The authors thank Rob Fender and the anonymous referee for valuable comments, which helped us to improve the paper substantially.
AAM thanks the UKRI Stephen Hawking fellowship.
The work of AYP was supported by the Ministry of Science and Higher Education of the Russian Federation (Agreement No. 075-15-2024-647). The work of VFS was supported by the German Research Foundation (DFG) grant WE\,1312/59-1. ST acknowledges support from the  Research Council of Finland Centre of Excellence in Neutron-Star Physics (grant 374064). This research was supported by the International Space Science Institute (ISSI) in Bern, through the International Team project 25-657 `Polarimetric Insights into Extreme Magnetism'.

\section*{Data availability}

The calculations presented in this paper were performed using a private code developed and owned by the corresponding author. 
All data appearing in the figures are available upon request. 

\bibliographystyle{mnras}
\bibliography{allbib}

@ARTICLE{2012A&A...542L..28K,
       author = {{Klochkov}, D. and {Doroshenko}, V. and {Santangelo}, A. and {Staubert}, R. and {Ferrigno}, C. and {Kretschmar}, P. and {Caballero}, I. and {Wilms}, J. and {Kreykenbohm}, I. and {Pottschmidt}, K. and {Rothschild}, R.~E. and {Wilson-Hodge}, C.~A. and {P{\"u}hlhofer}, G.},
        title = "{Outburst of GX 304-1 monitored with INTEGRAL: positive correlation between the cyclotron line energy and flux}",
      journal = {\aap},
         year = 2012,
        month = jun,
       volume = {542},
          eid = {L28},
        pages = {L28},
          doi = {10.1051/0004-6361/201219385},
archivePrefix = {arXiv},
       eprint = {1205.5475},
 primaryClass = {astro-ph.HE},
       adsurl = {https://ui.adsabs.harvard.edu/abs/2012A&A...542L..28K}
}

@ARTICLE{2007A&A...465L..25S,
       author = {{Staubert}, R. and {Shakura}, N.~I. and {Postnov}, K. and {Wilms}, J. and {Rothschild}, R.~E. and {Coburn}, W. and {Rodina}, L. and {Klochkov}, D.},
        title = "{Discovery of a flux-related change of the cyclotron line energy in Hercules X-1}",
      journal = {\aap},
         year = 2007,
        month = apr,
       volume = {465},
       number = {2},
        pages = {L25-L28},
          doi = {10.1051/0004-6361:20077098},
archivePrefix = {arXiv},
       eprint = {astro-ph/0702490},
 primaryClass = {astro-ph},
       adsurl = {https://ui.adsabs.harvard.edu/abs/2007A&A...465L..25S}
}

@ARTICLE{2002ApJ...580..389C,
       author = {{Campana}, S. and {Stella}, L. and {Israel}, G.~L. and {Moretti}, A. and {Parmar}, A.~N. and {Orlandini}, M.},
        title = "{The Quiescent X-Ray Emission of Three Transient X-Ray Pulsars}",
      journal = {\apj},
         year = 2002,
        month = nov,
       volume = {580},
       number = {1},
        pages = {389-393},
          doi = {10.1086/343074},
archivePrefix = {arXiv},
       eprint = {astro-ph/0207422},
 primaryClass = {astro-ph},
       adsurl = {https://ui.adsabs.harvard.edu/abs/2002ApJ...580..389C}
}

@ARTICLE{2024MNRAS.528.7320S,
       author = {{Shui}, Qing C. and {Zhang}, S. and {Wang}, Peng J. and {Mushtukov}, Alexander A. and {Santangelo}, A. and {Zhang}, Shuang N. and {Kong}, Ling D. and {Ji}, L. and {Chen}, Yu P. and {Doroshenko}, V. and {Frontera}, F. and {Chang}, Z. and {Peng}, Jing Q. and {Yin}, Hong X. and {Qu}, Jin L. and {Tao}, L. and {Ge}, Ming Y. and {Li}, J. and {Ye}, Wen T. and {Li}, Pan P.},
        title = "{Cyclotron line evolution revealed with pulse-to-pulse analysis in the 2020 outburst of 1A 0535+262}",
      journal = {\mnras},
         year = 2024,
        month = mar,
       volume = {528},
       number = {4},
        pages = {7320-7332},
          doi = {10.1093/mnras/stae352},
archivePrefix = {arXiv},
       eprint = {2403.11815},
 primaryClass = {astro-ph.HE},
       adsurl = {https://ui.adsabs.harvard.edu/abs/2024MNRAS.528.7320S}
}

@ARTICLE{2021ApJ...919...33C,
       author = {{Chen}, X. and {Wang}, W. and {Tang}, Y.~M. and {Ding}, Y.~Z. and {Tuo}, Y.~L. and {Mushtukov}, A.~A. and {Nishimura}, O. and {Zhang}, S.~N. and {Ge}, M.~Y. and {Song}, L.~M. and {Lu}, F.~J. and {Zhang}, S. and {Qu}, J.~L.},
        title = "{Relation of Cyclotron Resonant Energy and Luminosity in a Strongly Magnetized Neutron Star GRO J1008-57 Observed by Insight-HXMT}",
      journal = {\apj},
         year = 2021,
        month = sep,
       volume = {919},
       number = {1},
          eid = {33},
        pages = {33},
          doi = {10.3847/1538-4357/ac1268},
archivePrefix = {arXiv},
       eprint = {2107.03267},
 primaryClass = {astro-ph.HE},
       adsurl = {https://ui.adsabs.harvard.edu/abs/2021ApJ...919...33C}
}

@ARTICLE{2014PhyU...57..735P,
       author = {{Potekhin}, A. Y.},
        title = "{Atmospheres and radiating surfaces of neutron stars}",
      journal = {Physics Uspekhi},
         year = 2014,
        month = aug,
       volume = {57},
       number = {8},
          eid = {735-770},
        pages = {735-770},
          doi = {10.3367/UFNe.0184.201408a.0793},
archivePrefix = {arXiv},
       eprint = {1403.0074},
 primaryClass = {astro-ph.SR},
       adsurl = {https://ui.adsabs.harvard.edu/abs/2014PhyU...57..735P}
}

@ARTICLE{2024MNRAS.530..730M,
       author = {{Mushtukov}, A.~A. and {Ingram}, A. and {Suleimanov}, V.~F. and {DiLullo}, N. and {Middleton}, M. and {Tsygankov}, S.~S. and {van der Klis}, M. and {Portegies Zwart}, S.},
        title = "{Magnetospheric flows in X-ray pulsars - I. Instability at super-Eddington regime of accretion}",
      journal = {\mnras},
         year = 2024,
        month = may,
       volume = {530},
       number = {1},
        pages = {730-742},
          doi = {10.1093/mnras/stae781},
archivePrefix = {arXiv},
       eprint = {2402.12965},
 primaryClass = {astro-ph.HE},
       adsurl = {https://ui.adsabs.harvard.edu/abs/2024MNRAS.530..730M}
}

@ARTICLE{2019A&A...621A.134T,
       author = {{Tsygankov}, Sergey S. and {Doroshenko}, Victor and {Mushtukov}, Alexander A. and {Lutovinov}, Alexander A. and {Poutanen}, Juri},
        title = "{Study of the X-ray pulsar IGR J19294+1816 with NuSTAR: Detection of cyclotron line and transition to accretion from the cold disk}",
      journal = {\aap},
         year = 2019,
        month = jan,
       volume = {621},
          eid = {A134},
        pages = {A134},
          doi = {10.1051/0004-6361/201833786},
archivePrefix = {arXiv},
       eprint = {1811.08912},
 primaryClass = {astro-ph.HE},
       adsurl = {https://ui.adsabs.harvard.edu/abs/2019A&A...621A.134T}
}

@ARTICLE{2017A&A...608A..17T,
       author = {{Tsygankov}, S.~S. and {Mushtukov}, A.~A. and {Suleimanov}, V.~F. and {Doroshenko}, V. and {Abolmasov}, P.~K. and {Lutovinov}, A.~A. and {Poutanen}, J.},
        title = "{Stable accretion from a cold disc in highly magnetized neutron stars}",
      journal = {\aap},
         year = 2017,
        month = nov,
       volume = {608},
          eid = {A17},
        pages = {A17},
          doi = {10.1051/0004-6361/201630248},
archivePrefix = {arXiv},
       eprint = {1703.04528},
 primaryClass = {astro-ph.HE},
       adsurl = {https://ui.adsabs.harvard.edu/abs/2017A&A...608A..17T}
}

@ARTICLE{2019MNRAS.487L..30T,
       author = {{Tsygankov}, Sergey S. and {Doroshenko}, Victor and {Mushtukov}, Alexander A. and {Suleimanov}, Valery F. and {Lutovinov}, Alexander A. and {Poutanen}, Juri},
        title = "{Cyclotron emission, absorption, and the two faces of X-ray pulsar A 0535+262}",
      journal = {\mnras},
         year = 2019,
        month = jul,
       volume = {487},
       number = {1},
        pages = {L30-L34},
          doi = {10.1093/mnrasl/slz079},
archivePrefix = {arXiv},
       eprint = {1905.09496},
 primaryClass = {astro-ph.HE},
       adsurl = {https://ui.adsabs.harvard.edu/abs/2019MNRAS.487L..30T}
}

@ARTICLE{2021ApJ...912...17L,
       author = {{Lutovinov}, A. and {Tsygankov}, S. and {Molkov}, S. and {Doroshenko}, V. and {Mushtukov}, A. and {Arefiev}, V. and {Lapshov}, I. and {Tkachenko}, A. and {Pavlinsky}, M.},
        title = "{SRG/ART-XC and NuSTAR Observations of the X-Ray pulsar GRO J1008-57 in the Lowest Luminosity State}",
      journal = {\apj},
         year = 2021,
        month = may,
       volume = {912},
       number = {1},
          eid = {17},
        pages = {17},
          doi = {10.3847/1538-4357/abec43},
archivePrefix = {arXiv},
       eprint = {2103.05728},
 primaryClass = {astro-ph.HE},
       adsurl = {https://ui.adsabs.harvard.edu/abs/2021ApJ...912...17L}
}

@ARTICLE{2019MNRAS.483L.144T,
       author = {{Tsygankov}, Sergey S. and {Rouco Escorial}, Alicia and {Suleimanov}, Valery F. and {Mushtukov}, Alexander A. and {Doroshenko}, Victor and {Lutovinov}, Alexander A. and {Wijnands}, Rudy and {Poutanen}, Juri},
        title = "{Dramatic spectral transition of X-ray pulsar GX 304-1 in low luminous state}",
      journal = {\mnras},
         year = 2019,
        month = feb,
       volume = {483},
       number = {1},
        pages = {L144-L148},
          doi = {10.1093/mnrasl/sly236},
archivePrefix = {arXiv},
       eprint = {1810.13307},
 primaryClass = {astro-ph.HE},
       adsurl = {https://ui.adsabs.harvard.edu/abs/2019MNRAS.483L.144T}
}

@ARTICLE{2021AstBu..76....6F,
       author = {{Fabrika}, S.~N. and {Atapin}, K.~E. and {Vinokurov}, A.~S. and {Sholukhova}, O.~N.},
        title = "{Ultraluminous X-Ray Sources}",
      journal = {Astrophysical Bulletin},
         year = 2021,
        month = jan,
       volume = {76},
       number = {1},
        pages = {6-38},
          doi = {10.1134/S1990341321010077},
archivePrefix = {arXiv},
       eprint = {2105.10537},
 primaryClass = {astro-ph.GA},
       adsurl = {https://ui.adsabs.harvard.edu/abs/2021AstBu..76....6F}
}

@ARTICLE{1997ApJ...483..414P,
       author = {{Potekhin}, Alexander Y. and {Pavlov}, George G.},
        title = "{Photoionization of Hydrogen in Atmospheres of Magnetic Neutron Stars}",
      journal = {\apj},
         year = 1997,
        month = jul,
       volume = {483},
       number = {1},
        pages = {414-425},
          doi = {10.1086/304250},
archivePrefix = {arXiv},
       eprint = {astro-ph/9702004},
 primaryClass = {astro-ph},
       adsurl = {https://ui.adsabs.harvard.edu/abs/1997ApJ...483..414P}
}

@ARTICLE{1993ApJ...407..330P,
       author = {{Potekhin}, Aleksandr Y. and {Pavlov}, George G.},
        title = "{Photoionization of the Hydrogen Atom in Strong Magnetic Fields}",
      journal = {\apj},
         year = 1993,
        month = apr,
       volume = {407},
        pages = {330},
          doi = {10.1086/172515},
       adsurl = {https://ui.adsabs.harvard.edu/abs/1993ApJ...407..330P},
}

@ARTICLE{2015MNRAS.447.1847M,
       author = {{Mushtukov}, Alexander A. and {Suleimanov}, Valery F. and {Tsygankov}, Sergey S. and {Poutanen}, Juri},
        title = "{The critical accretion luminosity for magnetized neutron stars}",
      journal = {\mnras},
         year = 2015,
        month = feb,
       volume = {447},
       number = {2},
        pages = {1847-1856},
          doi = {10.1093/mnras/stu2484},
archivePrefix = {arXiv},
       eprint = {1409.6457},
 primaryClass = {astro-ph.HE},
       adsurl = {https://ui.adsabs.harvard.edu/abs/2015MNRAS.447.1847M}
}

@ARTICLE{2018SSRv..214...58B,
       author = {{Ballester}, Jos{\'e} Luis and {Alexeev}, Igor and {Collados}, Manuel and {Downes}, Turlough and {Pfaff}, Robert F. and {Gilbert}, Holly and {Khodachenko}, Maxim and {Khomenko}, Elena and {Shaikhislamov}, Ildar F. and {Soler}, Roberto and {V{\'a}zquez-Semadeni}, Enrique and {Zaqarashvili}, Teimuraz},
        title = "{Partially Ionized Plasmas in Astrophysics}",
      journal = {\ssr},
         year = 2018,
        month = mar,
       volume = {214},
       number = {2},
          eid = {58},
        pages = {58},
          doi = {10.1007/s11214-018-0485-6},
archivePrefix = {arXiv},
       eprint = {1707.07975},
 primaryClass = {astro-ph.SR},
       adsurl = {https://ui.adsabs.harvard.edu/abs/2018SSRv..214...58B}
}

@ARTICLE{2004ApJ...612.1034P,
       author = {{Potekhin}, Alexander Y. and {Lai}, Dong and {Chabrier}, Gilles and {Ho}, Wynn C.~G.},
        title = "{Electromagnetic Polarization in Partially Ionized Plasmas with Strong Magnetic Fields and Neutron Star Atmosphere Models}",
      journal = {\apj},
         year = 2004,
        month = sep,
       volume = {612},
       number = {2},
        pages = {1034-1043},
          doi = {10.1086/422679},
archivePrefix = {arXiv},
       eprint = {astro-ph/0405383},
 primaryClass = {astro-ph},
       adsurl = {https://ui.adsabs.harvard.edu/abs/2004ApJ...612.1034P}
}

@ARTICLE{2007ARep...51..549S,
       author = {{Suleimanov}, V.~F. and {Lipunova}, G.~V. and {Shakura}, N.~I.},
        title = "{The thickness of accretion {\ensuremath{\alpha}}-disks: Theory and observations}",
      journal = {Astronomy Reports},
         year = 2007,
        month = jul,
       volume = {51},
       number = {7},
        pages = {549-562},
          doi = {10.1134/S1063772907070049},
       adsurl = {https://ui.adsabs.harvard.edu/abs/2007ARep...51..549S}
}

@BOOK{1992pavi.book.....S,
       author = {{Shu}, F.~H.},
        title = "{The Physics of Astrophysics. Volume II: Gas Dynamics.}",
         year = 1992,
       adsurl = {https://ui.adsabs.harvard.edu/abs/1992pavi.book.....S},
publisher = "University Science Books",
      address = {Mill Valley, CA},
      isbn = {978-0935702651}
}

@ARTICLE{2015MNRAS.454.2714M,
       author = {{Mushtukov}, Alexander A. and {Tsygankov}, Sergey S. and {Serber}, Alexander V. and {Suleimanov}, Valery F. and {Poutanen}, Juri},
        title = "{Positive correlation between the cyclotron line energy and luminosity in sub-critical X-ray pulsars: Doppler effect in the accretion channel}",
      journal = {\mnras},
         year = 2015,
        month = dec,
       volume = {454},
       number = {3},
        pages = {2714-2721},
          doi = {10.1093/mnras/stv2182},
archivePrefix = {arXiv},
       eprint = {1509.05628},
 primaryClass = {astro-ph.HE},
       adsurl = {https://ui.adsabs.harvard.edu/abs/2015MNRAS.454.2714M}
}

@ARTICLE{2021A&A...651A..12S,
       author = {{Sokolova-Lapa}, E. and {Gornostaev}, M. and {Wilms}, J. and {Ballhausen}, R. and {Falkner}, S. and {Postnov}, K. and {Thalhammer}, P. and {F{\"u}rst}, F. and {Garc{\'\i}a}, J.~A. and {Shakura}, N. and {Becker}, P.~A. and {Wolff}, M.~T. and {Pottschmidt}, K. and {H{\"a}rer}, L. and {Malacaria}, C.},
        title = "{X-ray emission from magnetized neutron star atmospheres at low mass-accretion rates. I. Phase-averaged spectrum}",
      journal = {\aap},
         year = 2021,
        month = jul,
       volume = {651},
          eid = {A12},
        pages = {A12},
          doi = {10.1051/0004-6361/202040228},
archivePrefix = {arXiv},
       eprint = {2104.06802},
 primaryClass = {astro-ph.HE},
       adsurl = {https://ui.adsabs.harvard.edu/abs/2021A&A...651A..12S}
}

@ARTICLE{2021MNRAS.503.5193M,
       author = {{Mushtukov}, Alexander A. and {Suleimanov}, Valery F. and {Tsygankov}, Sergey S. and {Portegies Zwart}, Simon},
        title = "{Spectrum formation in X-ray pulsars at very low mass accretion rate: Monte Carlo approach}",
      journal = {\mnras},
         year = 2021,
        month = may,
       volume = {503},
       number = {4},
        pages = {5193-5203},
          doi = {10.1093/mnras/stab811},
archivePrefix = {arXiv},
       eprint = {2006.13596},
 primaryClass = {astro-ph.HE},
       adsurl = {https://ui.adsabs.harvard.edu/abs/2021MNRAS.503.5193M}
}

@ARTICLE{1973A&A....24..337S,
       author = {{Shakura}, N.~I. and {Sunyaev}, R.~A.},
        title = "{Black holes in binary systems. Observational appearance.}",
      journal = {\aap},
         year = 1973,
        month = jan,
       volume = {24},
        pages = {337-355},
       adsurl = {https://ui.adsabs.harvard.edu/abs/1973A&A....24..337S}
}

@BOOK{1986rpa..book.....R,
       author = {{Rybicki}, George B. and {Lightman}, Alan P.},
        title = "{Radiative Processes in Astrophysics}",
         year = 1979,
       publisher = {Wiley},
      address = {New York}
}

@ARTICLE{2022ApJ...941L..14T,
       author = {{Tsygankov}, Sergey S. and {Doroshenko}, Victor and {Poutanen}, Juri and {Heyl}, Jeremy and {Mushtukov}, Alexander A. and {Caiazzo}, Ilaria and {Di Marco}, Alessandro and {Forsblom}, Sofia V. and {Gonz{\'a}lez-Caniulef}, Denis and {Klawin}, Moritz and {La Monaca}, Fabio and {Malacaria}, Christian and {Marshall}, Herman L. and {Muleri}, Fabio and {Ng}, Mason and {Suleimanov}, Valery F. and {Sunyaev}, Rashid A. and {Turolla}, Roberto and {Agudo}, Iv{\'a}n and {Antonelli}, Lucio A. and {Bachetti}, Matteo and {Baldini}, Luca and {Baumgartner}, Wayne H. and {Bellazzini}, Ronaldo and {Bianchi}, Stefano and {Bongiorno}, Stephen D. and {Bonino}, Raffaella and {Brez}, Alessandro and {Bucciantini}, Niccol{\`o} and {Capitanio}, Fiamma and {Castellano}, Simone and {Cavazzuti}, Elisabetta and {Ciprini}, Stefano and {Costa}, Enrico and {De Rosa}, Alessandra and {Del Monte}, Ettore and {Di Gesu}, Laura and {Di Lalla}, Niccol{\`o} and {Donnarumma}, Immacolata and {Dov{\v{c}}iak}, Michal and {Ehlert}, Steven R. and {Enoto}, Teruaki and {Evangelista}, Yuri and {Fabiani}, Sergio and {Ferrazzoli}, Riccardo and {Garcia}, Javier A. and {Gunji}, Shuichi and {Hayashida}, Kiyoshi and {Iwakiri}, Wataru and {Jorstad}, Svetlana G. and {Karas}, Vladimir and {Kitaguchi}, Takao and {Kolodziejczak}, Jeffery J. and {Krawczynski}, Henric and {Latronico}, Luca and {Liodakis}, Ioannis and {Maldera}, Simone and {Manfreda}, Alberto and {Marin}, Fr{\'e}d{\'e}ric and {Marinucci}, Andrea and {Marscher}, Alan P. and {Matt}, Giorgio and {Mitsuishi}, Ikuyuki and {Mizuno}, Tsunefumi and {Ng}, Chi-Yung and {O'Dell}, Stephen L. and {Omodei}, Nicola and {Oppedisano}, Chiara and {Papitto}, Alessandro and {Pavlov}, George G. and {Peirson}, Abel L. and {Perri}, Matteo and {Pesce-Rollins}, Melissa and {Petrucci}, Pierre-Olivier and {Pilia}, Maura and {Possenti}, Andrea and {Puccetti}, Simonetta and {Ramsey}, Brian D. and {Rankin}, John and {Ratheesh}, Ajay and {Romani}, Roger W. and {Sgr{\`o}}, Carmelo and {Slane}, Patrick and {Soffitta}, Paolo and {Spandre}, Gloria and {Tamagawa}, Toru and {Tavecchio}, Fabrizio and {Taverna}, Roberto and {Tawara}, Yuzuru and {Tennant}, Allyn F. and {Thomas}, Nicholas E. and {Tombesi}, Francesco and {Trois}, Alessio and {Vink}, Jacco and {Weisskopf}, Martin C. and {Wu}, Kinwah and {Xie}, Fei and {Zane}, Silvia and {IXPE Collaboration}},
        title = "{The X-Ray Polarimetry View of the Accreting Pulsar Cen X-3}",
      journal = {\apjl},
         year = 2022,
        month = dec,
       volume = {941},
       number = {1},
          eid = {L14},
        pages = {L14},
          doi = {10.3847/2041-8213/aca486},
archivePrefix = {arXiv},
       eprint = {2209.02447},
 primaryClass = {astro-ph.HE},
       adsurl = {https://ui.adsabs.harvard.edu/abs/2022ApJ...941L..14T}
}

@ARTICLE{1976SvAL....2..111S,
       author = {{Sunyaev}, R.~A.},
        title = "{Plasma and radiation processes at the Alfven surface of an accreting neutron star.}",
      journal = {Soviet Astronomy Letters},
         year = 1976,
        month = jun,
       volume = {2},
        pages = {111-114},
       adsurl = {https://ui.adsabs.harvard.edu/abs/1976SvAL....2..111S}
}

@ARTICLE{2023MNRAS.525.4176B,
       author = {{Brice}, N. and {Zane}, S. and {Taverna}, R. and {Turolla}, R. and {Wu}, K.},
        title = "{Observational properties of accreting neutron stars with an optically thick envelope}",
      journal = {\mnras},
         year = 2023,
        month = nov,
       volume = {525},
       number = {3},
        pages = {4176-4185},
          doi = {10.1093/mnras/stad2391},
       adsurl = {https://ui.adsabs.harvard.edu/abs/2023MNRAS.525.4176B}
}

@ARTICLE{2019MNRAS.484..687M,
       author = {{Mushtukov}, Alexander A. and {Ingram}, Adam and {Middleton}, Matthew and {Nagirner}, Dmitrij I. and {van der Klis}, Michiel},
        title = "{Timing properties of ULX pulsars: optically thick envelopes and outflows}",
      journal = {\mnras},
         year = 2019,
        month = mar,
       volume = {484},
       number = {1},
        pages = {687-697},
          doi = {10.1093/mnras/sty3525},
archivePrefix = {arXiv},
       eprint = {1811.02049},
 primaryClass = {astro-ph.HE},
       adsurl = {https://ui.adsabs.harvard.edu/abs/2019MNRAS.484..687M}
}

@ARTICLE{2017MNRAS.467.1202M,
       author = {{Mushtukov}, Alexander A. and {Suleimanov}, Valery F. and {Tsygankov}, Sergey S. and {Ingram}, Adam},
        title = "{Optically thick envelopes around ULXs powered by accreating neutron stars}",
      journal = {\mnras},
         year = 2017,
        month = may,
       volume = {467},
       number = {1},
        pages = {1202-1208},
          doi = {10.1093/mnras/stx141},
archivePrefix = {arXiv},
       eprint = {1612.00964},
 primaryClass = {astro-ph.HE},
       adsurl = {https://ui.adsabs.harvard.edu/abs/2017MNRAS.467.1202M}
}

@ARTICLE{2015ARep...59..645S,
       author = {{Shakura}, N.~I. and {Postnov}, K.~A. and {Kochetkova}, A. Yu. and {Hjalmarsdotter}, L. and {Sidoli}, L. and {Paizis}, A.},
        title = "{Wind accretion: Theory and observations}",
      journal = {Astronomy Reports},
         year = 2015,
        month = jul,
       volume = {59},
       number = {7},
        pages = {645-655},
          doi = {10.1134/S1063772915070112},
archivePrefix = {arXiv},
       eprint = {1407.3163},
 primaryClass = {astro-ph.HE},
       adsurl = {https://ui.adsabs.harvard.edu/abs/2015ARep...59..645S}
}

@ARTICLE{2012MNRAS.420..216S,
       author = {{Shakura}, N. and {Postnov}, K. and {Kochetkova}, A. and {Hjalmarsdotter}, L.},
        title = "{Theory of quasi-spherical accretion in X-ray pulsars}",
      journal = {\mnras},
         year = 2012,
        month = feb,
       volume = {420},
       number = {1},
        pages = {216-236},
          doi = {10.1111/j.1365-2966.2011.20026.x},
archivePrefix = {arXiv},
       eprint = {1110.3701},
 primaryClass = {astro-ph.HE},
       adsurl = {https://ui.adsabs.harvard.edu/abs/2012MNRAS.420..216S}
}

@ARTICLE{2005AstL...31..729F,
       author = {{Filippova}, E.~V. and {Tsygankov}, S.~S. and {Lutovinov}, A.~A. and {Sunyaev}, R.~A.},
        title = "{Hard Spectra of X-ray Pulsars from INTEGRAL Data}",
      journal = {Astronomy Letters},
         year = 2005,
        month = nov,
       volume = {31},
       number = {11},
        pages = {729-747},
          doi = {10.1134/1.2123288},
archivePrefix = {arXiv},
       eprint = {astro-ph/0509525},
 primaryClass = {astro-ph},
       adsurl = {https://ui.adsabs.harvard.edu/abs/2005AstL...31..729F}
}

@ARTICLE{2009A&A...508..751S,
       author = {{Schure}, K.~M. and {Kosenko}, D. and {Kaastra}, J.~S. and {Keppens}, R. and {Vink}, J.},
        title = "{A new radiative cooling curve based on an up-to-date plasma emission code}",
      journal = {\aap},
         year = 2009,
        month = dec,
       volume = {508},
       number = {2},
        pages = {751-757},
          doi = {10.1051/0004-6361/200912495},
archivePrefix = {arXiv},
       eprint = {0909.5204},
 primaryClass = {astro-ph.GA},
       adsurl = {https://ui.adsabs.harvard.edu/abs/2009A&A...508..751S}
}

@ARTICLE{2025arXiv250909860X,
       author = {{Xiao}, Hua and {Tsygankov}, Sergey S. and {Suleimanov}, Valery F. and {Mushtukov}, Alexander A. and {Ji}, Long and {Poutanen}, Juri},
        title = "{Propeller effect in action: Unveiling quenched accretion in the transient X-ray pulsar 4U 0115+63}",
      journal = {arXiv e-prints},
         year = 2025,
        month = sep,
          eid = {arXiv:2509.09860},
        pages = {arXiv:2509.09860},
          doi = {10.48550/arXiv.2509.09860},
archivePrefix = {arXiv},
       eprint = {2509.09860},
 primaryClass = {astro-ph.HE},
       adsurl = {https://ui.adsabs.harvard.edu/abs/2025arXiv250909860X}
}

@ARTICLE{1980Ap&SS..73...33P,
       author = {{Pavlov}, G.~G. and {Shibanov}, Yu. A. and {Yakovlev}, D.~G.},
        title = "{Quantum Effects in Cyclotron Plasma Absorption}",
      journal = {\apss},
         year = 1980,
        month = nov,
       volume = {73},
       number = {1},
        pages = {33-62},
          doi = {10.1007/BF00642366},
       adsurl = {https://ui.adsabs.harvard.edu/abs/1980Ap&SS..73...33P}
}

@ARTICLE{2012ApJ...751...15S,
       author = {{Suleimanov}, V.~F. and {Pavlov}, G.~G. and {Werner}, K.},
        title = "{Magnetized Neutron Star Atmospheres: Beyond the Cold Plasma Approximation}",
      journal = {\apj},
         year = 2012,
        month = may,
       volume = {751},
       number = {1},
          eid = {15},
        pages = {15},
          doi = {10.1088/0004-637X/751/1/15},
archivePrefix = {arXiv},
       eprint = {1201.5527},
 primaryClass = {astro-ph.HE},
       adsurl = {https://ui.adsabs.harvard.edu/abs/2012ApJ...751...15S}
}

@BOOK{1995pprc.book.....D,
       author = {{Dolginov}, A.~Z. and {Gnedin}, Y.~N. and {Silant'ev}, N.~A.},
        title = "{Propagation and Polarisation of Radiation in Cosmic Media}",
         year = 1995,
    publisher = {Gordon and Breach Publ.},
      address = {Amsterdam},
       adsurl = {https://ui.adsabs.harvard.edu/abs/1995pprc.book.....D}
}

@ARTICLE{1975A&A....39..185I,
       author = {{Illarionov}, A.~F. and {Sunyaev}, R.~A.},
        title = "{Why the Number of Galactic X-ray Stars Is so Small?}",
      journal = {\aap},
         year = 1975,
        month = feb,
       volume = {39},
        pages = {185},
       adsurl = {https://ui.adsabs.harvard.edu/abs/1975A&A....39..185I}
}

@ARTICLE{2006ApJ...646..304U,
       author = {{Ustyugova}, G.~V. and {Koldoba}, A.~V. and {Romanova}, M.~M. and {Lovelace}, R.~V.~E.},
        title = "{``Propeller'' Regime of Disk Accretion to Rapidly Rotating Stars}",
      journal = {\apj},
         year = 2006,
        month = jul,
       volume = {646},
       number = {1},
        pages = {304-318},
          doi = {10.1086/503379},
archivePrefix = {arXiv},
       eprint = {astro-ph/0603249},
 primaryClass = {astro-ph},
       adsurl = {https://ui.adsabs.harvard.edu/abs/2006ApJ...646..304U}
}

@ARTICLE{2016A&A...593A..16T,
       author = {{Tsygankov}, S.~S. and {Lutovinov}, A.~A. and {Doroshenko}, V. and {Mushtukov}, A.~A. and {Suleimanov}, V. and {Poutanen}, J.},
        title = "{Propeller effect in two brightest transient X-ray pulsars: 4U 0115+63 and V 0332+53}",
      journal = {\aap},
         year = 2016,
        month = aug,
       volume = {593},
          eid = {A16},
        pages = {A16},
          doi = {10.1051/0004-6361/201628236},
archivePrefix = {arXiv},
       eprint = {1602.03177},
 primaryClass = {astro-ph.HE},
       adsurl = {https://ui.adsabs.harvard.edu/abs/2016A&A...593A..16T}
}

@ARTICLE{2019A&A...622A..61S,
       author = {{Staubert}, R. and {Tr{\"u}mper}, J. and {Kendziorra}, E. and {Klochkov}, D. and {Postnov}, K. and {Kretschmar}, P. and {Pottschmidt}, K. and {Haberl}, F. and {Rothschild}, R.~E. and {Santangelo}, A. and {Wilms}, J. and {Kreykenbohm}, I. and {F{\"u}rst}, F.},
        title = "{Cyclotron lines in highly magnetized neutron stars}",
      journal = {\aap},
         year = 2019,
        month = feb,
       volume = {622},
          eid = {A61},
        pages = {A61},
          doi = {10.1051/0004-6361/201834479},
archivePrefix = {arXiv},
       eprint = {1812.03461},
 primaryClass = {astro-ph.HE},
       adsurl = {https://ui.adsabs.harvard.edu/abs/2019A&A...622A..61S}
}

@ARTICLE{2023NewAR..9601672K,
       author = {{King}, Andrew and {Lasota}, Jean-Pierre and {Middleton}, Matthew},
        title = "{Ultraluminous X-ray sources}",
      journal = {\nar},
         year = 2023,
        month = jun,
       volume = {96},
          eid = {101672},
        pages = {101672},
          doi = {10.1016/j.newar.2022.101672},
archivePrefix = {arXiv},
       eprint = {2302.10605},
 primaryClass = {astro-ph.HE},
       adsurl = {https://ui.adsabs.harvard.edu/abs/2023NewAR..9601672K}
}

@BOOK{2002apa..book.....F,
       author = {{Frank}, Juhan and {King}, Andrew and {Raine}, Derek J.},
        title = "{Accretion Power in Astrophysics}",
publisher = "Cambridge University Press",
address = "Cambridge, UK",
         year = 2002,
edition = "3rd",
       adsurl = {https://ui.adsabs.harvard.edu/abs/2002apa..book.....F},
      isbn = {978-0521629577}
}

@INPROCEEDINGS{2022arXiv220414185M,
       author = {{Mushtukov}, Alexander and {Tsygankov}, Sergey},
        title = "{Accreting strongly magnetised neutron stars: X-ray pulsars}",
   booktitle    = {Handbook of X-ray and Gamma-ray Astrophysics},
  publisher    = {Springer},
  address      = {Singapore},
    editor     = {{Bambi}, Cosimo and {Santangelo}, Andrea},
     journal = {arXiv e-prints},
         year = 2024,
        month = apr,
          eid = {arXiv:2204.14185},
        pages = {4105-4136},
          doi = {10.1007/978-981-19-6960-7},
archivePrefix = {arXiv},
       eprint = {2204.14185},
 primaryClass = {astro-ph.HE},
       adsurl = {https://ui.adsabs.harvard.edu/abs/2022arXiv220414185M},
}

@ARTICLE{GinzburgOzernoi,
       author = {{Ginzburg}, Vitaliy L and {Ozernoi}, Lev M},
       title  = {On gravitational collapse of a magnetic stars},
      journal = {Sov. Phys. JETP},
         year = 1965,
       volume = 20,
        pages = {689}
}

@ARTICLE{PavlovMeszaros93,
       author = {{Pavlov}, G.~G. and {M\'esz\'aros}, P.},
        title = "{Finite-Velocity Effects on Atoms in Strong Magnetic Fields and Implications for Neutron Star Atmospheres}",
      journal = {\apj},
         year = 1993,
        month = oct,
       volume = {416},
        pages = {752},
          doi = {10.1086/173274},
       adsurl = {https://ui.adsabs.harvard.edu/abs/1993ApJ...416..752P}
}

@ARTICLE{VinckeBaye88,
       author = {{Vincke}, M. and {Baye}, D.},
        title = "{Centre-of-mass effects on the hydrogen atom in a magnetic field}",
      journal = {Journal of Physics B Atomic Molecular Physics},
         year = 1988,
        month = jul,
       volume = {21},
       number = {13},
        pages = {2407-2424},
          doi = {10.1088/0953-4075/21/13/009},
       adsurl = {https://ui.adsabs.harvard.edu/abs/1988JPhB...21.2407V}
}

@ARTICLE{Mitra98,
       author = {{Mitra}, Abhas},
        title = "{Maximum Accretion Efficiency in General Theory of Relativity}",
      journal = {arXiv e-prints},
         year = 1998,
        month = nov,
          eid = {astro-ph/9811402},
        pages = {astro-ph/9811402},
          doi = {10.48550/arXiv.astro-ph/9811402},
archivePrefix = {arXiv},
       eprint = {astro-ph/9811402},
 primaryClass = {astro-ph},
       adsurl = {https://ui.adsabs.harvard.edu/abs/1998astro.ph.11402M}
}

@ARTICLE{Meisel_18,
       author = {{Meisel}, Zach and {Deibel}, Alex and {Keek}, Laurens and {Shternin}, Peter and {Elfritz}, Justin},
        title = "{Nuclear physics of the outer layers of accreting neutron stars}",
      journal = {Journal of Physics G Nuclear Physics},
         year = 2018,
        month = sep,
       volume = {45},
       number = {9},
        pages = {093001},
          doi = {10.1088/1361-6471/aad171},
archivePrefix = {arXiv},
       eprint = {1807.01150},
 primaryClass = {astro-ph.HE},
       adsurl = {https://ui.adsabs.harvard.edu/abs/2018JPhG...45i3001M}
}

@ARTICLE{PotekhinCS99,
       author = {{Potekhin}, Alexander Y. and {Chabrier}, Gilles and {Shibanov}, Yuri A.},
        title = "{Partially ionized hydrogen plasma in strong magnetic fields}",
      journal = {\pre},
         year = 1999,
        month = aug,
       volume = {60},
       number = {2},
        pages = {2193-2208},
          doi = {10.1103/PhysRevE.60.2193},
archivePrefix = {arXiv},
       eprint = {astro-ph/9907006},
 primaryClass = {astro-ph},
       adsurl = {https://ui.adsabs.harvard.edu/abs/1999PhRvE..60.2193P}
}

@ARTICLE{Rogers00,
       author = {{Rogers}, Forrest J.},
        title = "{Ionization equilibrium and equation of state in strongly coupled plasmas}",
      journal = {Physics of Plasmas},
         year = 2000,
        month = jan,
       volume = {7},
       number = {1},
        pages = {51-58},
          doi = {10.1063/1.873815},
       adsurl = {https://ui.adsabs.harvard.edu/abs/2000PhPl....7...51R}
}

@ARTICLE{JohnsonHY83,
       author = {{Johnson}, Bruce R. and {Hirschfelder}, Joseph O. and {Yang}, Kuo-Ho},
        title = "{Interaction of atoms, molecules, and ions with constant electric and magnetic fields}",
      journal = {Reviews of Modern Physics},
         year = 1983,
        month = jan,
       volume = {55},
       number = {1},
        pages = {109-153},
          doi = {10.1103/RevModPhys.55.109},
       adsurl = {https://ui.adsabs.harvard.edu/abs/1983RvMP...55..109J}
}

@ARTICLE{HummerMihalas88,
       author = {{Hummer}, D.~G. and {Mihalas}, Dimitri},
        title = "{The Equation of State for Stellar Envelopes. I. an Occupation Probability Formalism for the Truncation of Internal Partition Functions}",
      journal = {\apj},
         year = 1988,
        month = aug,
       volume = {331},
        pages = {794},
          doi = {10.1086/166600},
       adsurl = {https://ui.adsabs.harvard.edu/abs/1988ApJ...331..794H}
}

@ARTICLE{InglisTeller39I,
       author = {{Inglis}, D.~R. and {Teller}, E.},
        title = "{Ionic Depression of Series Limits in One-Electron Spectra.}",
      journal = {\apj},
         year = 1939,
        month = oct,
       volume = {90},
        pages = {439},
          doi = {10.1086/144118},
       adsurl = {https://ui.adsabs.harvard.edu/abs/1939ApJ....90..439I}
}

@ARTICLE{DappenAM87,
       author = {{D{\"a}ppen}, Werner and {Anderson}, Lawrence and {Mihalas}, Dimitri},
        title = "{Statistical Mechanics of Partially Ionized Stellar Plasmas: The Planck-Larkin Partition Function, Polarization Shifts, and Simulations of Optical Spectra}",
      journal = {\apj},
         year = 1987,
        month = aug,
       volume = {319},
        pages = {195},
          doi = {10.1086/165446},
       adsurl = {https://ui.adsabs.harvard.edu/abs/1987ApJ...319..195D}
}

@ARTICLE{Potekhin96,
       author = {{Potekhin}, Alexander Y.},
        title = "{Ionization equilibrium of hot hydrogen plasma}",
      journal = {Physics of Plasmas},
         year = 1996,
        month = nov,
       volume = {3},
       number = {11},
        pages = {4156-4165},
          doi = {10.1063/1.871547},
archivePrefix = {arXiv},
       eprint = {plasm-ph/9607001},
 primaryClass = {physics.plasm-ph},
       adsurl = {https://ui.adsabs.harvard.edu/abs/1996PhPl....3.4156P}
}

@ARTICLE{PotekhinChabrier13,
       author = {{Potekhin}, A.~Y. and {Chabrier}, G.},
        title = "{Equation of state for magnetized Coulomb plasmas}",
      journal = {\aap},
         year = 2013,
        month = feb,
       volume = {550},
          eid = {A43},
        pages = {A43},
          doi = {10.1051/0004-6361/201220082},
archivePrefix = {arXiv},
       eprint = {1212.3405},
 primaryClass = {physics.plasm-ph},
       adsurl = {https://ui.adsabs.harvard.edu/abs/2013A&A...550A..43P}
}

@ARTICLE{Rosmej_20,
    author = {{Rosmej}, F.~B. and {Vainshtein}, L.~A. and {Astapenko}, V.~A. and {Lisitsa}, V.~S.},
        title = "{Statistical and quantum photoionization cross
sections in plasmas: Analytical approaches for any
conﬁgurations including inner shells}",
      journal = {Matter Radiat. Extremes},
         year = 2020,
        month = nov,
       volume = {5},
          eid = {064202},
          doi = {10.1063/5.0022751}
}

@ARTICLE{PavlovShibanov79,
       author = {{Pavlov}, G.~G. and {Shibanov}, Yu. A.},
        title = "{Influence of vacuum polarization by a magnetic field on the propagation of electromagnetic waves in a plasma}",
      journal = {Sov. Phys. -- JETP},
         year = 1979,
        month = may,
       volume = {49},
        pages = {741-749},
       adsurl = {https://ui.adsabs.harvard.edu/abs/1979JETP...49..741P},
}

@BOOK{Spitzer_book,
       author = {{Spitzer}, L.},
        title = "{Physics of Fully Ionized Gases}",
         year = 1962,
       adsurl = {https://ui.adsabs.harvard.edu/abs/1962pfig.book.....S},
publisher = "Interscience",
      address = {New York},
      isbn = {978-0470817230}
}

@ARTICLE{HardingLai06,
       author = {{Harding}, Alice K. and {Lai}, Dong},
        title = "{Physics of strongly magnetized neutron stars}",
      journal = {Reports on Progress in Physics},
         year = 2006,
        month = sep,
       volume = {69},
       number = {9},
        pages = {2631-2708},
          doi = {10.1088/0034-4885/69/9/R03},
archivePrefix = {arXiv},
       eprint = {astro-ph/0606674},
 primaryClass = {astro-ph},
       adsurl = {https://ui.adsabs.harvard.edu/abs/2006RPPh...69.2631H}
}

@ARTICLE{MushtukovNP16,
       author = {{Mushtukov}, Alexander A. and {Nagirner}, Dmitrij I. and {Poutanen}, Juri},
        title = "{Compton scattering S matrix and cross section in strong magnetic field}",
      journal = {\prd},
         year = 2016,
        month = may,
       volume = {93},
       number = {10},
          eid = {105003},
        pages = {105003},
          doi = {10.1103/PhysRevD.93.105003},
archivePrefix = {arXiv},
       eprint = {1512.06681},
 primaryClass = {astro-ph.HE},
       adsurl = {https://ui.adsabs.harvard.edu/abs/2016PhRvD..93j5003M}
}

@BOOK{Meszaros_book,
       author = {{M\'esz\'aros}, Peter},
        title = "{High-energy radiation from magnetized neutron stars}",
       address= "{Chicago}",
  publisher   = "{University of Chicago Press}",
         year = 1992,
         isbn = {978-0226520940}
}

@ARTICLE{PotekhinLai07,
       author = {{Potekhin}, Alexander Y. and {Lai}, Dong},
        title = "{Statistical equilibrium and ion cyclotron absorption/emission in strongly magnetized plasmas}",
      journal = {\mnras},
         year = 2007,
        month = apr,
       volume = {376},
       number = {2},
        pages = {793-808},
          doi = {10.1111/j.1365-2966.2007.11474.x},
archivePrefix = {arXiv},
       eprint = {astro-ph/0701285},
 primaryClass = {astro-ph},
       adsurl = {https://ui.adsabs.harvard.edu/abs/2007MNRAS.376..793P}
}

@ARTICLE{Ventura79,
       author = {{Ventura}, J.},
        title = "{Scattering of light in a strongly magnetized plasma}",
      journal = {\prd},
         year = 1979,
        month = mar,
       volume = {19},
       number = {6},
        pages = {1684-1695},
          doi = {10.1103/PhysRevD.19.1684},
       adsurl = {https://ui.adsabs.harvard.edu/abs/1979PhRvD..19.1684V}
}

@BOOK{Ginzburg70,
       author = {{Ginzburg}, V.~L.},
        title = "{The propagation of electromagnetic waves in plasmas}",
      edition = "{2nd}",
    publisher = "{Freeman and Co.}",
      address = {New York},
         year = 1970,
       adsurl = {https://ui.adsabs.harvard.edu/abs/1970pewp.book.....G},
}

@ARTICLE{1983A&A...118...66N,
       author = {{Nagel}, W. and {Ventura}, J.},
        title = "{Coulomb bremsstrahlung and cyclotron emissivity in hot magnetized plasmas}",
      journal = {\aap},
         year = 1983,
        month = feb,
       volume = {118},
       number = {1},
        pages = {66-74},
       adsurl = {https://ui.adsabs.harvard.edu/abs/1983A&A...118...66N}
}

@ARTICLE{1980ApJ...236..904N,
       author = {{Nagel}, W.},
        title = "{Cyclotron line formation in the accretion column of an X-ray pulsar}",
      journal = {\apj},
         year = 1980,
        month = mar,
       volume = {236},
        pages = {904-910},
          doi = {10.1086/157817},
       adsurl = {https://ui.adsabs.harvard.edu/abs/1980ApJ...236..904N}
}

@ARTICLE{1985ApJ...298..147M,
       author = {{Meszaros}, P. and {Nagel}, W.},
        title = "{X-ray pulsar models. I. Angle-dependent cyclotron line formation and comptonization.}",
      journal = {\apj},
         year = 1985,
        month = nov,
       volume = {298},
        pages = {147-160},
          doi = {10.1086/163594},
       adsurl = {https://ui.adsabs.harvard.edu/abs/1985ApJ...298..147M}
}

@ARTICLE{KadomtsevKudryavtsev71,
       author = {{Kadomtsev}, B.~B. and {Kudryavtsev}, V.~S.},
        title = "{Atoms in a superstrong magnetic field}",
      journal = {Sov. Phys. JETP Lett.},
         year = 1971,
        month = jan,
       volume = {13},
        pages = {42},
       adsurl = {https://ui.adsabs.harvard.edu/abs/1971JETPL..13...42K}
}

@ARTICLE{CohenLR70,
       author = {{Cohen}, R. and {Lodenquai}, J. and {Ruderman}, M.},
        title = "{Atoms in superstrong magnetic fields}",
      journal = {\prl},
         year = 1970,
        month = aug,
       volume = {25},
       number = {7},
        pages = {467-469},
          doi = {10.1103/PhysRevLett.25.467},
       adsurl = {https://ui.adsabs.harvard.edu/abs/1970PhRvL..25..467C}
}

@ARTICLE{Khersonskii87,
       author = {{Khersonskii}, V.~K.},
        title = "{Ionization equilibrium of atomic hydrogen in a strong magnetic field}",
      journal = {\sovast},
         year = 1987,
        month = apr,
       volume = {31},
        pages = {225},
       adsurl = {https://ui.adsabs.harvard.edu/abs/1987SvA....31..225K}
}

@ARTICLE{2010PhRvE..81b1126C,
       author = {{Chac{\'o}n-Acosta}, Guillermo and {Dagdug}, Leonardo and {Morales-T{\'e}cotl}, Hugo A.},
        title = "{Manifestly covariant J{\"u}ttner distribution and equipartition theorem}",
      journal = {\pre},
         year = 2010,
        month = feb,
       volume = {81},
       number = {2},
          eid = {021126},
        pages = {021126},
          doi = {10.1103/PhysRevE.81.021126},
archivePrefix = {arXiv},
       eprint = {0910.1625},
 primaryClass = {cond-mat.stat-mech},
       adsurl = {https://ui.adsabs.harvard.edu/abs/2010PhRvE..81b1126C}
}

@ARTICLE{KadomtsevKudryavtsev72,
       author = {{Kadomtsev}, B.~B. and {Kudryavtsev}, V.~S.},
        title = "{Matter in a Superstrong Magnetic Field}",
      journal = {Sov. Phys. JETP},
         year = 1972,
        month = jan,
       volume = {35},
        pages = {76},
       adsurl = {https://ui.adsabs.harvard.edu/abs/1972JETP...35...76K}
}

\appendix

\section{Photoionization of ground-state hydrogen atom in a strong magnetic field}
\label{app:PhotoIonCS}

\subsection{Photoionization cross section}

{High-frequency radiation in a strongly magnetized plasmas can be generally (except at resonances) described in terms of two normal modes: extraordinary (indexed by $j=1$) and ordinary ($j=2$) \citep{Ginzburg70}. In the semi-transverse approximation \citep{Ventura79}, 
their complex polarization vectors $\bm{e}_j$ are orthogonal to the wave vector and to each other.
It is convenient to use the cyclic basis vectors $\hat{\mathbf{e}}_{\pm1} = (\hat{\mathbf{e}}_x \pm \ii \hat{\mathbf{e}}_y)/\sqrt{2}$ and $\hat{\mathbf{e}}_0 = \hat{\mathbf{e}}_z$, where $\hat{\mathbf{e}}_{x,y,z}$ are the unit vectors along the  Cartesian axes with the $z$-axis directed along $\bm{B}$.
Then a cross section for interaction of a photon with polarization $j$, having energy $E$ and propagating in a direction $\bm{n}$, can be approximately} written as
\beq 
\sigma_\mathrm{ph}(j,E,\bm{n}) = \sum\limits_{\mu=-1}^{+1}
\sigma_\mathrm{ph}^{(\mu)}(E)
\left|e_{j,\mu}(E,\theta_B)\right|^2,
\label{sigmasum}
\eeq 
where $\theta_B$ is the angle of $\bm{n}$ to $\bm{B}$ and
$e_{j,\mu}(E,\theta_B) = \bm{e}_j(E,\theta_B) \cdot \hat{\mathbf{e}}_\mu^*$ are the cyclic components of the polarization vector. 
This representation is exact in the dipole approximation, which excludes the terms of the form 
$\sigma_\mathrm{ph}^{(\mu\mu')}(E) e_{j,\mu}(E,\theta_B) e_{j,\mu'}^*(E,\theta_B)$ with $\mu'\neq\mu$ for transitions between the states with definite $\bm{B}$-projections $-\hbar s$ of the orbital momentum \citep[see, e.g.,][]{1993ApJ...407..330P}. 

In this paper, for estimates, we limit ourselves by consideration of photoionization from hydrogen ground state only (i.e., with the discrete quantum numbers $s=\nu=0$) and neglect the effects of thermal atomic motion across the magnetic field on the cross sections. Besides, we are interested in photon energies considerably below the cyclotron energy, at which the free electron remains on the ground Landau level.
In this case the photoionization with $\mu=-1$ is forbidden due to the angular momentum conservation, 
and it is sufficient to keep only terms with $\mu=0$ and $\mu=+1$ in the sum (\ref{sigmasum}).
The threshold energy for the photoionization of a non-moving H atom at $B \gg B_0$ equals
\beq 
E_\mathrm{th}^{(\mu)} = E_\mathrm{b} + \mu \left(\frac{m_\mathrm{e}}{m_\mathrm{p}}\right)E_\mathrm{cyc},
\eeq 
where $E_\mathrm{b}$ is the binding energy
(note that one should put $\sigma_\mathrm{ph}^{(\mu)}(E)=0$ at $E < E_\mathrm{th}^{(\mu)}$
in equation~(\ref{sigmasum})).
Practical approximations for the binding energies of H atoms in magnetic fields are given in \citet{2014PhyU...57..735P}. In particular, for the ground state of a non-moving H atom
\beq 
\frac{E_\mathrm{b}}{\mbox{Ry}} = 
\frac{ 1 + x/a_1 + a_3 x^3 + a_4 x^4 + a_6 x^6 }
{1+a_2 x^2 + a_5 x^3 + a_6 x^4}.
\label{E0s0}
\eeq 
where 
\[
x = \ln(1+a_1 B/B_0),
\]
$a_1 = 0.862
$, $
a_2 = 0.4513
$, $
a_3 = 0.2775
$, $
a_4 = 0.0557,
$, $
a_5 = 0.0431,
$ and $
a_6 = 0.002075.
$
Equation (\ref{E0s0}) is valid for any magnetic field strength.

A fit for the cross sections of photoionization of a non-moving H atom in fields $B \gg 100\,B_0$
has been constructed in \citet{1993ApJ...407..330P}. For the ground state it can be written in the form
\beq 
\frac{\sigma_\mathrm{ph}^{(\mu)}(E)}{\sigma_\mathrm{ph}^{(\mu)}(E_\mathrm{th}^{(\mu)})} =  \left(\frac{E}{E_\mathrm{th}^{(\mu)}}\right)^{\!\!|\mu|}
\!\! \frac{1}{(1+a_\mu y)^{5/2}}
\frac{1}{[1+b_\mu(\sqrt{1+y}-1)]^4},
\eeq 
where $E > E_\mathrm{th}^{(\mu)}$, $y = (E - E_\mathrm{th}^{(\mu)})/E_\mathrm{b}$,
$a_0 = 1.15$,  $a_{+1} = 0.72$, $b_0 = 0.84(B_0/B)^{2/7}$ and
$b_{+1} = 0.91(B_0/B)^{2/5}$.

The cross sections at the threshold energies is given for each polarization
($\mu=0$ or $\mu=+1$) by the fits
\begin{align}&
\frac{\sigma_\mathrm{ph}^{(0)}(E_\mathrm{th}^{(0)})}{\sigma_\mathrm{T}} =
\frac{5.14\times 10^6\,\sqrt{\mbox{Ry}/E_\mathrm{b}}}{
1 + 0.0234\,[ \ln(B/B_0) - 4.51 \,]^2},
\\&
\frac{\sigma_\mathrm{ph}^{(+1)}(E_\mathrm{th}^{(+1)})}{\sigma_\mathrm{T}} = 
\frac{E_\mathrm{th}^{(+1)}}{E_\mathrm{b}}\frac{B_0}{B}
\frac{9\times 10^5\,\sqrt{\mbox{Ry}/E_\mathrm{b}}}{
1+0.0141\ln^2(B/(2B_0))},
\end{align}
where we have neglected small angle-dependent factors, which should be zero in the consistent dipole approximation.
If $E + E_\mathrm{b} < E_\mathrm{cyc} - m_\mathrm{e}/m_\mathrm{p}$, which is the case we consider in this paper, then, as mentioned above, $\sigma_\mathrm{ph}^{(-1)}(E) =0$. This selection rule does not hold for moving atoms, but $\sigma_\mathrm{ph}^{(-1)}$ is very small for small pseudomomenta $K_\perp$ \citep[see][]{1997ApJ...483..414P}.

Note that the approximations reproduced in this appendix are valid for atoms at rest.
They are sufficiently accurate for slowly moving atoms in tightly bound states.
For thermally distributed atoms, the formalism of calculations of the photoionization cross sections has been described by \citet{1997ApJ...483..414P}. The validity range and accuracy of the approximations for the non-moving atoms can be evaluated using the perturbation theory \citep{VinckeBaye88,PavlovMeszaros93}. 
A simple order-of-magnitude
evaluation of its applicability, which was suggested in \citet{2014PhyU...57..735P}, reduces for the hydrogen atom to $E_\mathrm{b} \gg T B_{12}/4$,
where $E_\mathrm{b}$ is given by equation~(\ref{E0s0}). When $B_{12}$ {increases from 5 to 100}, the ground-state energy $E_\mathrm{b}$ increases from {258 eV to 534 eV}. Therefore the approximations for the cross sections in this appendix are good for the ground state at $T$ {below a few tens eV, otherwise they can be used only as crude order-of-magnitude estimates}. Since binding energies of excited states are smaller, the applicability ranges for them will be narrower.

\subsection{Photoionization and recombination rates at equilibrium}
\label{app:rates}

\begin{figure}
\centering 
\includegraphics[width=\columnwidth]{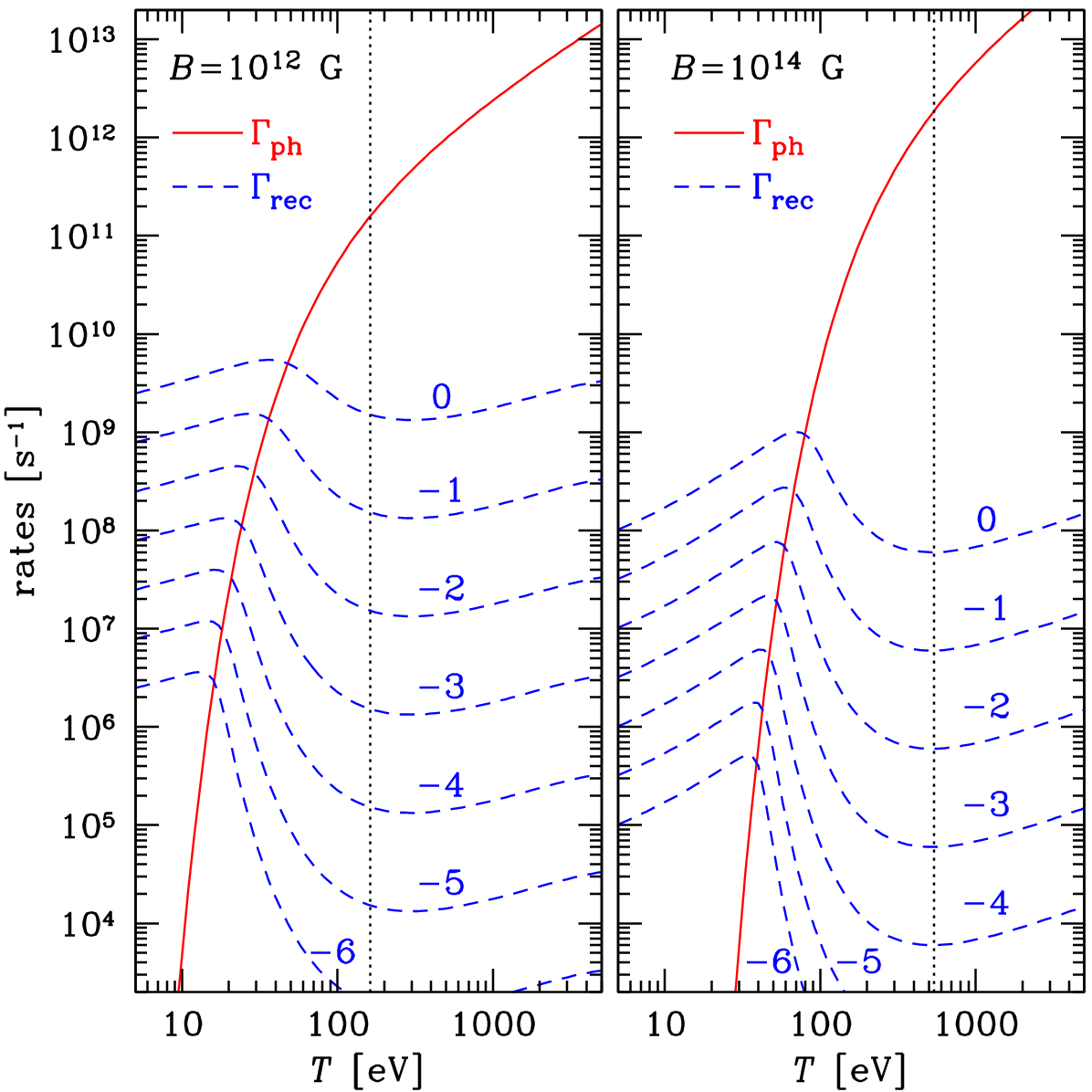}
\caption{
{Equilibrium photoionization (solid lines) and recombination (dashed lines) rates as functions of temperature at $B=10^{12}$~G
(left panel) and $10^{14}$~G (right panel). Numbers at the dashed curves indicate the mass density logarithms, $\log_{10}\rho_1$. The vertical dotted line is drawn at $T=E_\mathrm{b}$, where the binding energy $E_\mathrm{b}$ is given by equation~(\ref{E0s0}).}
}
\label{pic:rates}
\end{figure}

{
In complete thermal equilibrium characterized by temperature $T$, the energy-related specific intensity (that is the radiative energy per unit photon energy $E$, per unit area of the surface orthogonal to $\bm{n}$ and per steradian) in each of the two normal modes $j=1,2$ equals one half of the Planck function
\beq
I_E(j,E,\bm{n}) = \frac{\mathcal{B}_E}{2}
= 
\,\frac{E^3}{8\pi^3 \hbar^3 c^2} 
\,\frac{1}{\ee^{E/T}-1}.
\eeq
The number of photons with energies between $E$ and $E+\dd E$ in a solid angle element $\dd\bm{n}$ crossing a surface element $\dd S$ perpendicular to $\bm{n}$ during a time interval $\dd t$ equals $(I_E(j,E,\bm{n})/E)\,\dd E \,\dd S \,\dd\bm{n} \,\dd t$. 
This photon flux ionizes
an atom with cross sections $\sigma_\mathrm{ph}(j,E,\bm{n})$ at a rate $\sigma_\mathrm{ph}(j,E,\bm{n})(I_E(j,E,\bm{n})/E)\,\dd E \,\dd\bm{n}$.
For both polarizations, integration over all photon energies and directions gives the photoionization rate
\beq
\Gamma_\mathrm{ph} = \sum_{j=1,2} \int_{(4\pi)}\dd\bm{n} \int\limits_0^\infty \,\dd E
\,\frac{\sigma_\mathrm{ph}(j,E,\bm{n})}{8\pi^3 \hbar^3 c^2} 
\,\frac{E^2}{\ee^{E/T}-1}.
\eeq
Using equation~(\ref{sigmasum}) and equality \citep{Ventura79}
\beq
\sum_{j=1,2} \int_{(4\pi)} |e_{j,\mu}(E,\theta_B)|^2 \,\dd\bm{n} = \frac{8\pi}{3},
\eeq
we obtain
\beq
\Gamma_\mathrm{ph} = \sum_{\mu=-1}^1 \int_{E_\mathrm{th}^{(\mu)}}^\infty \frac{\sigma_\mathrm{ph}^{(\mu)}(E)}{3\pi^2 \hbar^3 c^2} 
\,\frac{E^2}{\ee^{E/T}-1}\,\dd E.
\label{Gamma_ph}
\eeq
Since the cross section component $\sigma_\mathrm{ph}^{(0)}(E)$ dominates by orders of magnitude at $B\gg B_0$, one can safely retain only the term $\mu=0$ in equation~(\ref{Gamma_ph}). The photoionization rates (\ref{Gamma_ph}) as functions of temperature at $B=5\times10^{12}$~G and $10^{14}$~G are shown by solid lines in Fig.~\ref{pic:rates}.
}

{
The principle of detailed balance gives the recombination rate 
\beq
   \Gamma_\mathrm{rec} = \frac{n_\mathrm{H}}{n_\mathrm{e}}\,\Gamma_\mathrm{ph}
    = \frac{1-f_+}{f_+}\,\Gamma_\mathrm{ph}.
\label{Gamma_rec}
\eeq
{For estimates, let us again consider the ground-state atoms without motion effects in a non-degenerate hydrogen plasma, neglect the difference between $\lambda_\mathrm{p}$ and $\lambda_\mathrm{H}$, and discard the terms $F_\mathrm{ex,ep}$, $F_\mathrm{ex,b}$ and $F_\mathrm{mol}$ in equation~(\ref{Fren}).
Then the minimum of the free energy $F$ is delivered by
\beq
   n_\mathrm{H} = \frac{n_\mathrm{p}n_\mathrm{e}}{n_\mathrm{p,tot}}\,K(T),
\label{Saha_simple}
\eeq
where
\beq
   K(T) = \lambda_\mathrm{e}^3\, n_\mathrm{p,tot} 
                \,\frac{1-\ee^{-E_\mathrm{cyc}/T}}{E_\mathrm{cyc}/T}
                \,\frac{1-\ee^{-E_\mathrm{cyc,p}/T}}{E_\mathrm{cyc,p}/T}
                \,\ee^{E_\mathrm{b}/T}
\label{KT1}
\eeq
is the dimensionless equilibrium constant.
If $E_\mathrm{cyc,p} \ll T \ll E_\mathrm{cyc}$, then
}
\begin{align}&
   K(T) \approx 2\pi a_\mathrm{m}^2 \lambda_\mathrm{e}n_\mathrm{p,tot}\ee^{E_\mathrm{b}/T}
   \approx
   \frac{2.86\times10^{-8}}{B_{12}\sqrt{T_{100}}}\,n_{19}\,\ee^{E_\mathrm{b}/T}
\nonumber\\&\qquad
   \approx \frac{0.00171\,\rho_1}{B_{12}\sqrt{T_{100}}}\,\ee^{E_\mathrm{b}/T},
\label{KT}
\end{align}
where
$n_{19} \equiv n_\mathrm{p,tot}/10^{19}\text{~cm}^{-3}$
and $\rho_1 \equiv \rho/\text{1~g~cm}^{-3}$. 
}

Equation~(\ref{Saha_simple}) yields the ionized fraction in the explicit form
\beq
   f_+ = \frac{\sqrt{1+4K(T)}-1}{2K(T)}.
\label{approxSaha}
\eeq
A comparison with numerical solutions of equation~(\ref{magSaha}) shows that this approximation {provides an accuracy of $f_+$ within a factor of $\approx2$ at $B \leq 10^{14}$~G, being more accurate at weaker fields, although it gives an incorrect low-density asymptote for $f_\mathrm{H} = 1-f_+$ in the high-ionization ($f_\mathrm{H} \lesssim 0.1$) domain due to a large fraction of so called decentred states, brought about by atomic motion across $\bm{B}$ (neglected here) and surviving in a plasma at low densities 
\citep[see Figs.~6, 7 in][]{PotekhinCS99}. 
The recombination rates (\ref{Gamma_rec}) for the ionization degree $f_+$ given by equations~(\ref{approxSaha}) and (\ref{KT}) are shown as functions of temperature by dashed lines in Fig.~\ref{pic:rates} at $B=5\times10^{12}$~G and $10^{14}$~G for several densities from $\rho=10^{-6}$ g~cm$^{-3}$ ($n_\mathrm{p,tot}\approx6\times10^{17}$ cm$^{-3}$) to $\rho = 1$ g~cm$^{-3}$ ($n_\mathrm{p,tot}\approx6\times10^{23}$ cm$^{-3}$).
}

\section{Opacity in electron cyclotron line and its harmonics}
\label{app:CyclotronOpacity}

We compute the opacities due to electron cyclotron absorption and its harmonics following the approach developed by \citet*{1980Ap&SS..73...33P} \citep*[see also][]{2012ApJ...751...15S}. 
The cyclotron opacity is evaluated separately for each harmonic using semi-analytic expressions appropriate for a fully ionized hydrogen plasma in LTE.

The model assumes that the radiation propagates through a magnetized plasma characterized by a magnetic field $B$, temperature $T$, electron density 
$n_\mathrm{e}$ and angle $\theta_B$ between the photon propagation direction and the magnetic field. 
The photon energy $E$
determines the resonance condition for a given harmonic $s$, corresponding to 
$E \approx s\,E_\mathrm{cyc}$. 
All expressions below are valid in the non-relativistic limit, assuming $T \ll m_\mathrm{e} c^2$, $E \gg E_\mathrm{pl}$ and $E_\mathrm{cyc} \gg E_\mathrm{pl}$, where $E_\mathrm{pl} =  (4\pi\hbar^2 n_\mathrm{e} e^2 / m_\mathrm{e})^{1/2}
\approx
0.12\, \sqrt{n_\mathrm{e} / 10^{25}\mbox{~cm}^{-3}}$ keV is the characteristic plasma energy associated with the Langmuir frequency.

The opacity for a given normal mode $j$ and harmonic $s$ can be expressed as \citep[see][]{1980Ap&SS..73...33P,2012ApJ...751...15S}
\begin{equation}
    \kappa_j
=
    \left(
\frac{E_\mathrm{pl}}{E}
\right)^2 A_T (X_T \pm Y_T) \exp(-x_s^2),
\end{equation}
where
$x_s$ is a dimensionless detuning parameter describing the offset from the $s$-th cyclotron harmonic. 
{The upper sign corresponds to the extraordinary mode, and the lower to the ordinary mode.}
The terms $A_T$, $X_T$, and $Y_T$ account for Doppler broadening, thermal motion and polarization effects of the magnetized plasma:
\begin{align}
X_T &= 1 + \cos^{2}\theta_B + \zeta_T \sin^{2}\theta_B, 
\\
Y_T &= (1 + G_T^{2})^{-1/2}
   \left[ 2|\cos\theta_B| + G_T(1 - \zeta_T)\sin^{2}\theta_B \right],
\\
G_T &= \frac{\sin^{2}\theta_B}{2|\cos\theta_B|}
     \frac{E_\mathrm{cyc}^{4} + V_B E^{2}(E^{2} - E_\mathrm{cyc}^{2})}{E_\mathrm{cyc}^{3}E}, 
\\
\zeta_T &= \frac{\xi_T^{2}\beta_T^{2}\sin^{2}\theta_B}{4s b_T}
        \tanh\!\left(\frac{b_T}{2}\right), 
\\
A_T &= \frac{1}{8}
  \left(\frac{b_T}{1 - \ee^{-b_T}}\right)^{s}
     \frac{1 - \ee^{-\xi_T}}{\xi_T}\,
     \frac{\xi_T^{s}}{(s-1)!}\,
     \left( \frac{\beta_T E \sin\theta_B}{2E_\mathrm{cyc}} \right)^{2s-2}
\nonumber\\&\times
          \sqrt{\pi}\,\beta_T |\cos\theta_B|.
\end{align}
{Here $V_B$ is the vacuum polarization parameter, 
which at $b\equiv B/B_\mathrm{QED} \ll 1$ can be written as \citep{PavlovShibanov79}}
\beq
    V_B = \frac{3\times 10^{28}\mbox{~cm}^{-3}}{n_\mathrm{e}}\,b^4,
\eeq
$\beta_T = \sqrt{2T / m_\mathrm{e} c^2}$ is the mean thermal velocity of electrons in units of speed of light,
         $\xi_T = E/T$
and $b_T = E_\mathrm{cyc}/T$.

For each harmonic $s$, the photon energy is either input directly or, for numerical integration over angle and frequency, calculated using the resonance condition with relativistic corrections:
\begin{equation}
    E = \frac{s b_T - s^2 b_T^2 \beta_T^2 / 4}{T^{-1} - {0.25\,|\cos\theta_B|}/{\sqrt{
         m_\mathrm{e}c^2
\, T}}},
\end{equation}

The exponential term $\exp(-x_s^2)$ describes Doppler broadening of the resonance line profile, where
\begin{equation}
x_s = \frac{E - sE_\mathrm{cyc}}{\Delta E_D}, \qquad 
\Delta E_D = E_\mathrm{cyc}\,\beta_T\,|\cos\theta_B|
\end{equation}
is the Doppler width of the line.

\section{Magnetic free-free cooling in the strongly quantizing regime}
\label{sec:Magnetic_free-free_cooling_new}

\begin{figure*}
\centering 
\includegraphics[width=17cm]{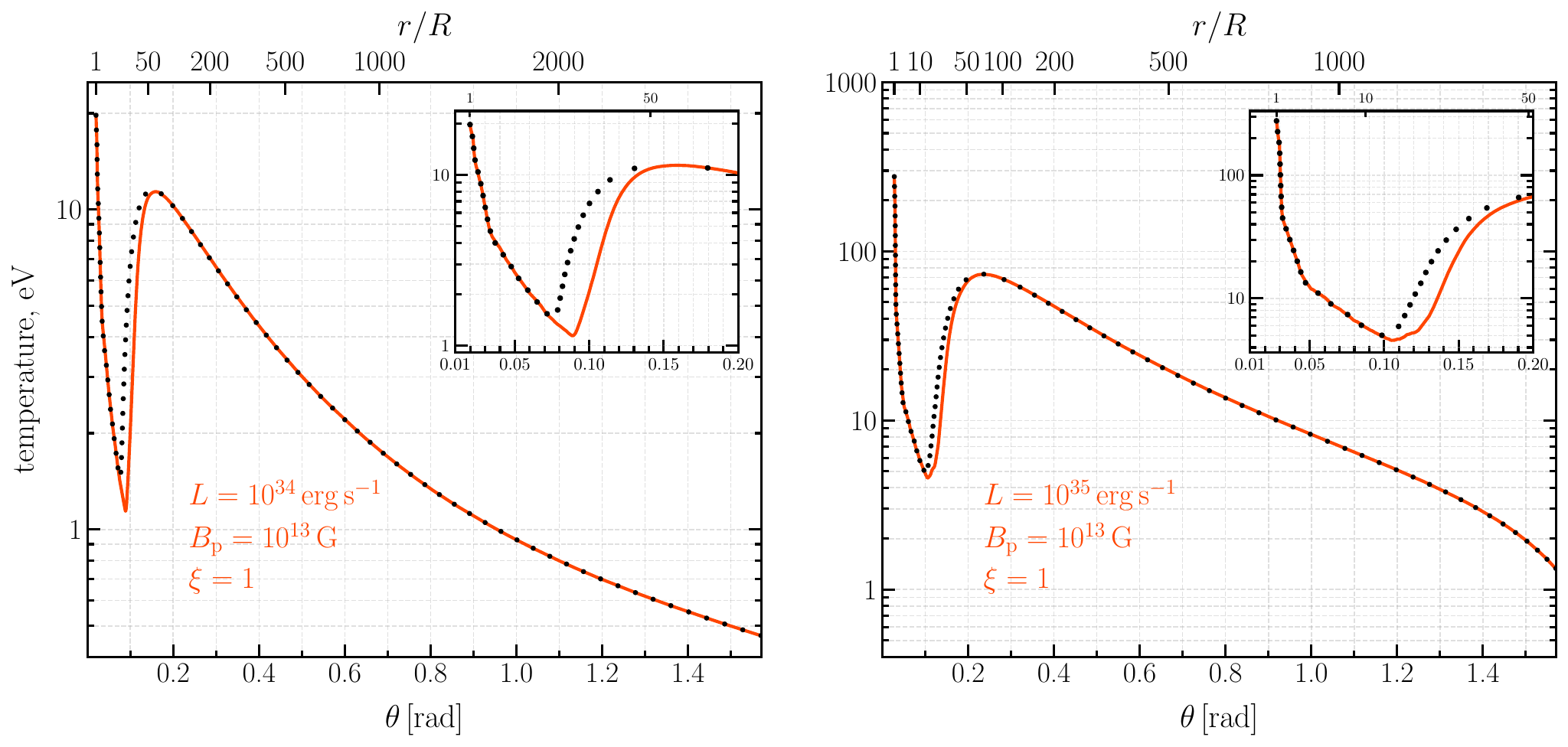}
\caption{
Sensitivity of the temperature profile to the treatment of cyclotron cooling.
The two panels show calculations for different accretion luminosities:
\(L=10^{34}\,{\rm erg\,s^{-1}}\) in the left panel and
\(L=10^{35}\,{\rm erg\,s^{-1}}\) in the right panel.
In both cases, the surface magnetic field is \(B_{\rm p}=10^{13}\,{\rm G}\) and the beaming parameter is \(\xi=1\).
The solid red curves show the fiducial calculations, while the black points show the limiting cases in which the cyclotron cooling term \(Q^{-}_{\rm cyc}\) is artificially removed from the thermal-balance calculation.
The upper horizontal axes show the radial distance from the neutron-star centre in units of the stellar radius, \(r/R\), while the lower horizontal axes show the magnetic colatitude \(\theta\).
The insets zoom in on the inner part of the accretion channel, close to the NS surface.
Removing the cyclotron cooling term changes the temperature mainly in a limited intermediate range of magnetic colatitudes, while the temperature in the near-surface region, where the ionization balance and photoionized layer are evaluated, remains practically unchanged.
This demonstrates that the near-surface temperature, and thus the conclusions of the paper on ionization, are insensitive to the uncertain LTE treatment of cyclotron emission.
}
\label{pic:sc_flow_temp_v7}
\end{figure*}

In the strongly magnetized part of the accretion channel, the treatment of cyclotron cooling requires special care. 
A direct LTE treatment of cyclotron emission implicitly assumes a Boltzmann population of the excited electron Landau levels. 
However, in a strongly quantizing magnetic field, the populations of these levels are controlled by collisional excitation, radiative de-excitation, and, in the presence of external photons near the cyclotron frequency, resonant absorption followed by re-emission.
Under the conditions typical of the lower accretion channel, radiative de-excitation is faster than collisional excitation, so the excited Landau levels are substantially underpopulated relative to the Boltzmann distribution,
\begin{equation}
\label{eq:landau_underpopulation}
    n_N \ll n_N^{\rm LTE}
    \propto
    \exp\left(-N E_{\rm cyc}/kT\right),
    \qquad N\geq 1 .
\end{equation}
A direct LTE treatment of cyclotron emission can therefore overestimate the true photon production rate.

The physically consistent treatment of resonant cyclotron cooling in this regime should include the cyclotron resonances as part of the magnetic free-free, or bremsstrahlung, process rather than as an independent LTE cyclotron-emission channel. 
This problem has been discussed in detail in the context of radiative transfer in strongly magnetized accretion columns by \citet{1980ApJ...236..904N}, \citet{1983A&A...118...66N}, and \citet{1985ApJ...298..147M}. 
A fully self-consistent implementation would require the calculation of magnetic free-free emissivities, including the resonant structure and the relevant polarization-dependent Gaunt factors, over the broad range of magnetic-field strengths and photon energies encountered along the accretion flow. 
Such a calculation is beyond the scope of the present paper.

We therefore use a simpler sensitivity test to assess whether the uncertain treatment of cyclotron cooling can affect our main conclusions. 
Since the approximate LTE cyclotron prescription used in Section~\ref{sec:free-free_emission} is expected to provide an upper-limit estimate to the true cyclotron photon production rate in the strongly quantizing regime, we repeated the temperature calculation after artificially setting \(Q^{-}_{\rm cyc}=0\). 
This limiting case removes the potentially overestimated cooling channel altogether and therefore provides a conservative test of the role of cyclotron cooling in the thermal balance.

The result is shown in Fig.~\ref{pic:sc_flow_temp_v7}. 
The removal of cyclotron cooling affects the temperature mainly in a limited intermediate region of the flow. 
In contrast, the near-surface temperature, which determines the recombination and photoionization calculations discussed in Sections~\ref{sec:Ionization_degree} and~\ref{sec:Ionisation_degree_above_NS}, remains almost unchanged. 
At larger distances the two solutions also converge, because the magnetic field is weaker and cyclotron cooling is inefficient. 
We therefore conclude that the partial recombination near the NS surface is controlled primarily by free-free cooling, Compton heating, and compressional heating, and is not an artefact of the approximate cyclotron cooling prescription.

\section{Effect of resonant magnetic scattering on the thermal balance}
\label{app:Resonant_scattering}

\begin{figure}
\centering 
\includegraphics[width=\columnwidth]{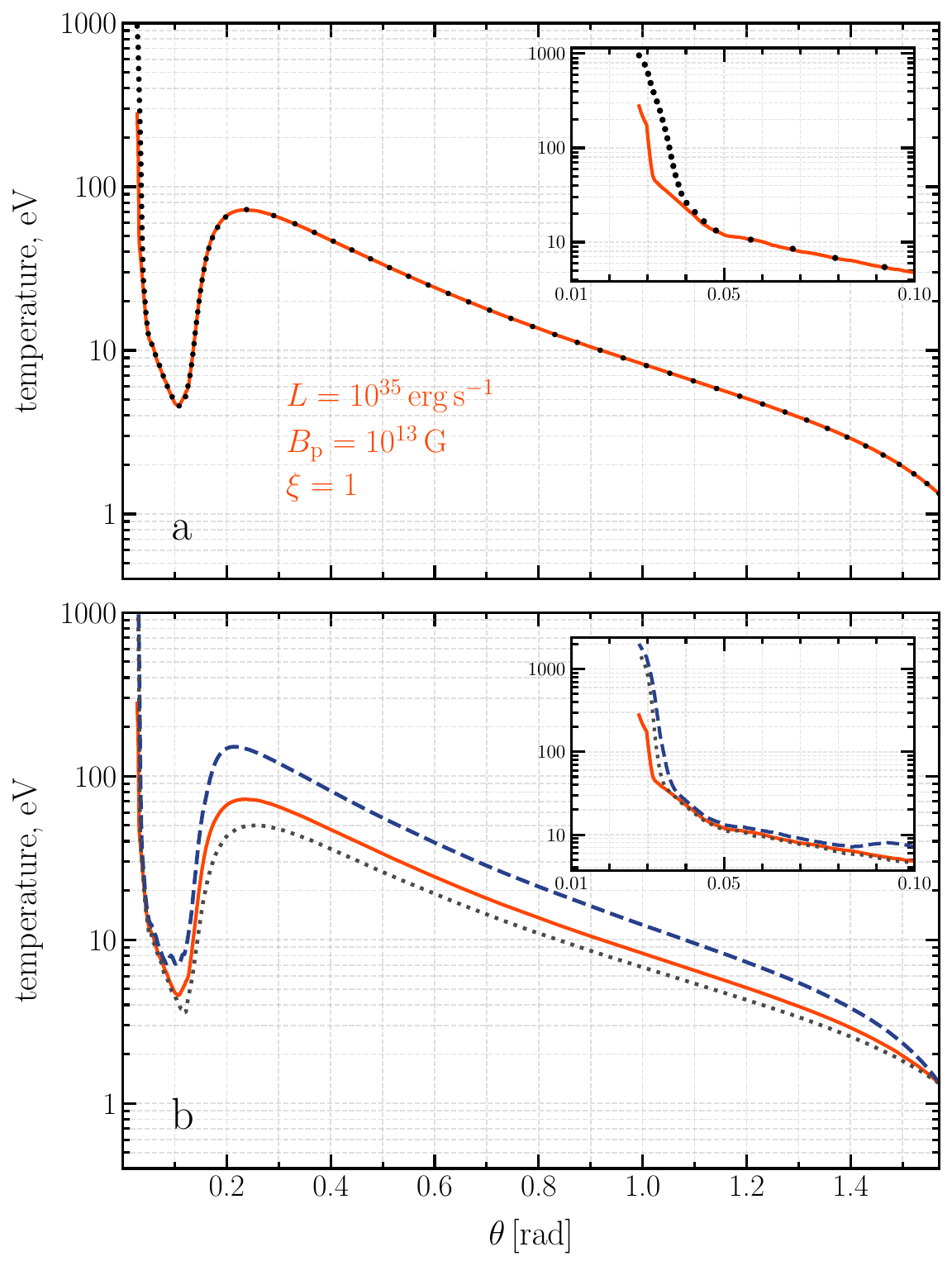}
\caption{
Temperature profiles of the accretion flow. 
Panel (a): Comparison of the temperature profiles of the accretion flow calculated without (solid line) and with (dotted line) inclusion of resonant magnetic scattering. 
The resonant scattering increases the temperature in the vicinity of the NS surface, while at larger distances the solutions converge. 
Panel (b) illustrates the effect of the spectral shape of the incident radiation. 
The red solid line corresponds to the blackbody case. 
The dashed line shows the result for a cutoff power-law spectrum with $(\Gamma, E_\mathrm{cut}) = (0.5, 15~\mathrm{keV})$, while the dotted line corresponds to a softer spectrum with $(\Gamma, E_\mathrm{cut}) = (1.0, 10~\mathrm{keV})$. 
The cutoff power-law spectra lead to higher temperatures in the inner region of the magnetosphere, whereas at larger distances the temperature may be either higher or lower compared to the blackbody case depending on the spectral parameters.
}
\label{pic:sc_flow_temp_v6}
\end{figure}

In the main text, Compton scattering is treated in the non-magnetic approximation. 
However, in the strong-field regime (at $r \lesssim 10R$), where the cyclotron energy cannot be considered as small with respect to the typical energies of photons emitted from the NS surface, photon scattering on the electrons can be strongly affected by the cyclotron resonance and depends on photon polarization state.
To estimate the possible impact of this effect on the thermal structure of the accretion flow, we performed calculations including the resonant scattering in a simplified manner.

{ 
To estimate the scattering cross sections in the electron rest frame, we apply the expressions  derived by \citet{Ventura79} in the non-relativistic approximation.
In the considered case of relatively low accretion luminosities, we can neglect the ellipticity of normal-mode polarization vectors and treat them as linearly polarized (see Section~\ref{sec:weak_strong}). 
In this approximation, the cross sections $\sigma_\mathrm{X}$ and $\sigma_\mathrm{O}$ for the X- and O-modes are given by
\begin{align}&
\frac{\sigma_\mathrm{X}}{\sigma_\mathrm{T}}
\approx
\frac12\left[
\frac{E^2}{(E-E_\mathrm{cyc})^2}
+
\frac{E^2}{(E+E_\mathrm{cyc})^2}
\right],
\\&
\frac{\sigma_\mathrm{O}}{\sigma_\mathrm{T}}
\approx
\sin^2\theta_B
+
\frac{\cos^2\theta_B}{2}\left[
\frac{E^2}{(E-E_\mathrm{cyc})^2}
+
\frac{E^2}{(E+E_\mathrm{cyc})^2}
\right].
\end{align}
}

{
In the vicinity of the resonance, 
following \citet{Ventura79},
we replace the divergent term by the Lorentzian approximation with a finite width:
\begin{equation}
\frac{E^2}{(E-E_\mathrm{cyc})^2}
\longrightarrow
\frac{E^2}{(E-E_\mathrm{cyc})^2+\Gamma_E^2/4},
\end{equation}
where
\begin{equation}
\Gamma_E = \frac43\, \frac{e^2 E^2}{\hbar m_\mathrm{e} c^3} 
\approx
\frac{4\alpha_\mathrm{f}}{3}\, \frac{E_\mathrm{cyc}^2}{m_\mathrm{e} c^2}
= 
\frac{4\alpha_\mathrm{f}}{3} \, \frac{B}{B_\mathrm{QED}\, E_\mathrm{cyc}}.
\end{equation}
is the natural radiative width of the first Landau level.
}

The inclusion of resonant scattering leads to an increase in the local heating rate in the inner region of the accretion flow, and the temperature near the NS surface becomes higher than in the non-magnetic case.
However, the effect is spatially localized. At distances exceeding a few stellar radii, the temperature profiles obtained with and without resonant scattering converge. 
The maximal increase of the temperature near the surface does not exceed a factor of $\sim 2$ for the considered parameter range.

The modest impact of resonant scattering can be attributed to several factors. 
The resonance condition $E \simeq E_\mathrm{cyc}(r)$ is satisfied only within a narrow spatial region due to the strong radial dependence of the magnetic field ($B \propto r^{-3}$). 
In addition, the solid angle under which the emitting surface is seen from a given point in the flow is small, which limits the flux of photons interacting resonantly.

Given these uncertainties, as well as the moderate quantitative effect on the thermal structure, we do not include resonant scattering in the main model. 
The comparison of temperature profiles with and without this effect is shown in Fig.~\ref{pic:sc_flow_temp_v6}a.

\section{Dependence on the spectral shape of the incident radiation} 
\label{app:Spectra}

In the main text, we model the radiation field produced at the NS surface assuming a blackbody spectrum. 
However, observed spectra of X-ray pulsars are typically described by harder continua, most commonly approximated by a power law with an exponential cutoff \citep{2005AstL...31..729F}. 
Such spectra are characterized by an extended high-energy tail and therefore potentially stronger Compton heating of accretion flow.

To estimate the sensitivity of our results to the spectral shape, we repeated the calculations adopting a cutoff power-law spectrum of the form
\begin{equation}
\frac{\dd N}{\dd E} = A\,E^{-\Gamma}\exp\left(-\frac{E}{E_\mathrm{cut}}\right),
\end{equation}
where $\Gamma$ is the photon index and $E_\mathrm{cut}$ is the cutoff energy. 
The normalization constant $A$ was chosen such that the bolometric luminosity is the same as in the blackbody case.
We consider two representative parameter sets, motivated by typical spectral fits of sub-critical XRPs at luminosities $L \sim 10^{35}$--$10^{36}\,\mathrm{erg\,s^{-1}}$:
\begin{equation}
(\Gamma, E_\mathrm{cut}) = (0.5,\,15~\mathrm{keV}) \quad \text{and} \quad (1.0,\,10~\mathrm{keV}).
\end{equation}
The first set corresponds to a relatively hard spectrum, while the second represents a softer continuum. 
We note that at very low luminosities the spectral shape may deviate from this form and become more complex (e.g. two-component spectra), but the adopted approximation is sufficient for our purposes.

The resulting temperature profiles are shown in Fig.~\ref{pic:sc_flow_temp_v6}b. 
At large distances from a NS, the temperature obtained with the cutoff power-law spectra can be either higher or lower than in the blackbody case, depending on the spectral parameters. 
In contrast, in the inner region of the magnetosphere, close to the NS surface, the cutoff power-law spectra systematically lead to higher temperatures compared to the blackbody case. 
Nevertheless, the overall effect remains moderate. 
For the considered parameter range, the temperature increase near the surface does not exceed a factor of $\sim 2$, and the global structure of the temperature profile is preserved.
We conclude that while the spectral shape of the incident radiation influences the heating rate in the inner magnetosphere, it does not qualitatively change the thermal structure of the accretion flow or the main conclusions of this work.

\bsp 
\label{lastpage}
\end{document}